\documentclass[acmsmall]{acmart}
\AtBeginDocument{%
  }

\setcopyright{acmlicensed}
\copyrightyear{2025}
\acmYear{2025}

\acmVolume{99}
\acmNumber{99}
\acmArticle{99}
\acmMonth{9}
\usepackage{enumitem}
\usepackage{multirow}
\usepackage{xcolor}
\usepackage{graphicx}      
\usepackage{subcaption}    
\usepackage{algorithm}
\usepackage{algorithmic}
\usepackage{amsmath}

\usepackage{amssymb}

\begin{document}

\title{Towards Multi-Behavior Multi-Task Recommendation via Behavior-informed Graph Embedding Learning}


\author{Wenhao Lai}
\email{laiwenhao2023@email.szu.edu.cn}

\affiliation{%
	\institution{College of Computer Science and Software Engineering, Shenzhen University}
	\streetaddress{3688\# Nanhai Avenue, Nanshan District}
	\city{Shenzhen}
	\postcode{518060}
	\country{China}
}

\author{Weike Pan}
\email{panweike@szu.edu.cn}
\affiliation{%
	\institution{College of Computer Science and Software Engineering, Shenzhen University}
	\streetaddress{3688\# Nanhai Avenue, Nanshan District}
	\city{Shenzhen}
	\postcode{518060}
	\country{China}
}

\author{Zhong Ming}
\email{mingz@szu.edu.cn}
\affiliation{%
	\institution{Guangdong Laboratory of Artificial Intelligence and Digital Economy (SZ)}
	\city{Shenzhen}
	\postcode{518123}
	\country{China}
}

\renewcommand{\shortauthors}{Lai et al.}

\begin{abstract}
  Multi-behavior recommendation (MBR) aims to improve the performance w.r.t. the target behavior (i.e., purchase) by leveraging auxiliary behaviors (e.g., click, favourite). However, in real-world scenarios, a recommendation method often needs to process different types of behaviors and generate personalized lists for each task (i.e., each behavior type). Such a new recommendation problem is referred to as multi-behavior multi-task recommendation (MMR). So far, the most powerful MBR methods usually model multi-behavior interactions using a cascading graph paradigm. Although significant progress has been made in optimizing the performance of the target behavior, it often neglects the performance of auxiliary behaviors. To compensate for the deficiencies of the cascading paradigm, we propose a novel solution for MMR, i.e., behavior-informed graph embedding learning (BiGEL). Specifically, we first obtain a set of behavior-aware embeddings by using a cascading graph paradigm. Subsequently, we introduce three key modules to improve the performance of the model. The cascading gated feedback (CGF) module enables a feedback-driven optimization process by integrating feedback from the target behavior to refine the auxiliary behaviors preferences. The global context enhancement (GCE) module integrates the global context to maintain the user's overall preferences, preventing the loss of key preferences due to individual behavior graph modeling. Finally, the contrastive preference alignment (CPA) module addresses the potential changes in user preferences during the cascading process by aligning the preferences of the target behaviors with the global preferences through contrastive learning. Extensive experiments on two real-world datasets demonstrate the effectiveness of our BiGEL compared with ten very competitive methods.
\end{abstract}




\maketitle

\section{Introduction}

Recommender systems (RS) are an important part of modern online platforms that provide personalized recommendations to users by analyzing their historical interactions. They play a key role in alleviating the problem of information overload, which is becoming increasingly prevalent in the digital age. The goal of these systems is to help users discover content or products that match their preferences and needs, ultimately improving user experience and business success. One of the most widely adopted approaches in recommender systems is collaborative filtering (CF)~\cite{gsurvey, gsurvey2}, which learns user preferences based on their interactions with items. CF models typically rely on user item interaction data to infer user interests, and the models usually focus on a single type of behavior such as click or purchase~\cite{NGCF, LightGCN, lightgt, BIGCF}.

However, with the development of online platforms, the interactions between users and items become more diverse. For example, on e-commerce platforms, users interact with items through multiple behaviors, such as click, favourite and purchase. When these interactions are considered individually, only a limited view of user preferences can be learned because each behavior type conveys only partial information about the user's interests. Traditional recommendation models based on one single behavior type have difficulty in capturing the full range of user needs, leading to suboptimal recommendations. To address this limitation, multi-behavior recommendation (MBR) has emerged as an effective solution.

Multi-behavior recommendation (MBR) extends the concept of single-behavior recommender systems to incorporate multiple types of user-item interactions. In MBR, one behavior is usually selected as the target behavior (e.g., purchase), while other behaviors, including click and favourite, are regarded as auxiliary behaviors. Auxiliary behaviors are then leveraged to improve the recommendation performance of the target behavior, providing a richer and more comprehensive understanding of user preferences. Early studies in MBR employed various sampling strategies to prioritize the target behavior, aiming to improve the prediction accuracy by focusing more on the target behavior while still leveraging auxiliary behaviors~\cite{VALS, Rdsm, MFBPR, BPRH}. These methods help overcome the limitations such as data sparsity by incorporating additional forms of interaction.

Graph convolutional networks (GCNs) are widely used in recommender systems for their ability to capture higher-order interaction information between users and items~\cite{survey1,lightgnn,robust,graphcl}. GCNs have been effectively used to model complex relationships in recommender systems, especially in situations in which interactions between users and items can be represented by graph structures of the scenarios. Some studies have naturally extended GCNs to multi-behavior scenarios by representing each type of user-item interaction as a different graph. In this case, user and item embeddings are learned from each behavior graph and complex dependencies between behaviors are captured through different mechanisms~\cite{VCGAE, MBSSL, HPMR, MULE}. This graph-based approach allows modeling complex relationships between different types of behaviors, considering not only individual interactions but also their interdependencies, thus improving the overall recommendation performance. Other studies have focused on modeling real-world scenarios where user interactions follow an ordered sequence of behaviors (e.g., $click \succ favourite \succ purchase$) and use a cascading graph paradigm to model dependencies between behaviors, where the information propagates unidirectionally from auxiliary behaviors to the target behavior in order to optimize the recommendation performance w.r.t. the target behavior~\cite{CRGCN, MBCGCN, AutoDCS, DAGCN}. The cascading graph paradigm effectively improves the accuracy of target behavior prediction and improves the overall performance of the recommender system by capturing the dependencies between behaviors.

\begin{figure*}[tbp]
    \centering 
    \includegraphics[width=0.8\linewidth]{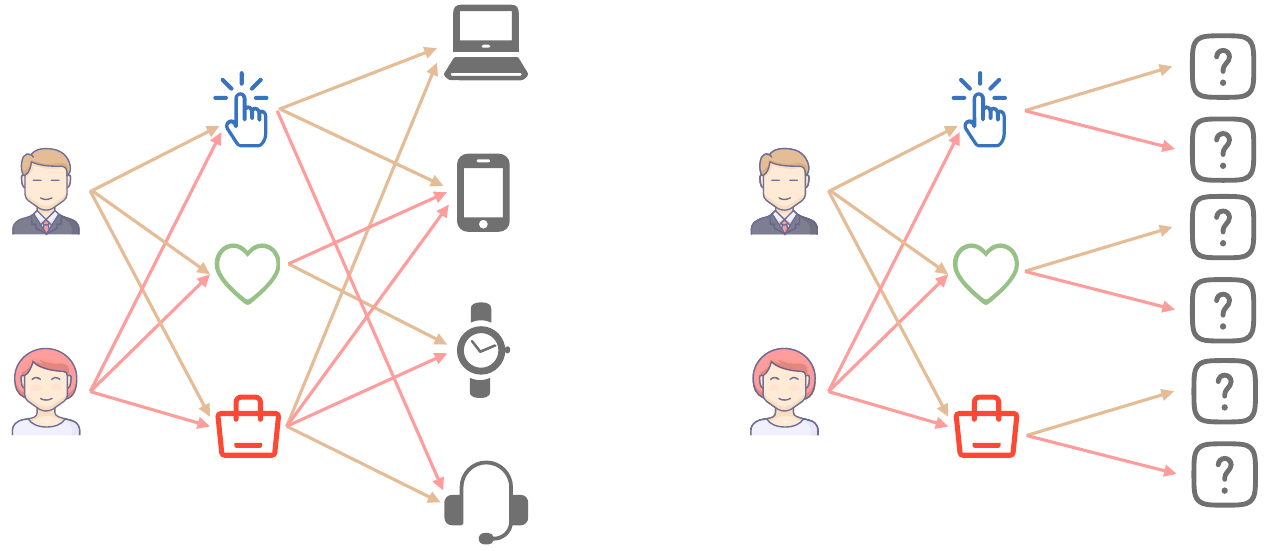} 
    \caption{Illustration of multi-behavior multi-task recommendation.} 
    \label{fig:intro} 
\end{figure*}

However, despite the success of the cascading graph paradigm in optimizing the target behavior, it tends to neglect the optimization of the auxiliary behaviors. This leads to an imbalance in which the auxiliary behaviors are used alone as signals to improve the performance w.r.t. the target behavior, but their own prediction accuracy is not directly improved. In real-world recommender systems, although the auxiliary behaviors (e.g., click, favourite) may not directly bring economic benefits, they play a crucial role in enhancing user engagement and promoting long-term retention~\cite{POGCN}. For example, the frequency of item click is a key indicator of platform activity. More attention to products also boosts seller engagement. Therefore, a real-world recommender system is often expected to be able to handle multiple behavior feedbacks and generate multi-task personalized recommendations w.r.t. these behaviors. This is known as multi-behavior multi-task recommendation (MMR). Figure~\ref{fig:intro} shows an example of multi-behavior multi-task recommendation, where users interact with items through three different behaviors, i.e., click, favourite, and purchase. The goal of multi-behavior multi-task recommendation is to recommend items for each user that are most likely to interact with under each behavior.

Recently, BVAE~\cite{BVAE} addresses the MMR problem by extending VAE~\cite{VAE}, while POGCN~\cite{POGCN} tackles it by extending LightGCN~\cite{LightGCN} to a partial order graph. Although these works have made some progress, they still fail to adequately capture the interdependencies between different behaviors. In contrast, the cascading graph paradigm has received much attention for its ability to capture the real interaction dependencies between user behaviors. This motivates us to explore how to extend the cascading graph paradigm to better address the MMR problem. In particular, we address the following two challenges in MMR:

\begin{itemize}
\item \textbf{Neglecting the recommendation performance w.r.t. the auxiliary behaviors.}
Most existing multi-behavior recommendation methods, such as VCGAE~\cite{VCGAE}, BCIPM~\cite{BCIPM}, CRGCN~\cite{CRGCN}, MBCGCN~\cite{MBCGCN} and DA-GCN~\cite{DAGCN}, predominantly focus on improving the recommendation performance w.r.t. the target behavior only. Although some methods~\cite{CRGCN, PKEF, DAGCN} incorporate multi-task learning by integrating some auxiliary behaviors with the target behavior, they do not explicitly model the relationships between different tasks, but treat the target behavior as the primary objective.

\item \textbf{User preference bias induced by cascading propagation.}
In the cascading graph paradigm, as it propagates from one behavior (e.g., click) to the next, if certain intermediate behaviors (e.g., favourite) are missing, the convolution operation may propagate the interaction information of other users who interact with the same items. Replacing the missing user behaviors with other users' interaction information will distort the preference representation and indirectly affect the subsequent behavior layers, leading to user preference bias.
\end{itemize}

To address the above two challenges, we propose a novel solution for MMR, i.e., behavior-informed graph embedding learning (BiGEL), which extends the cascading graph paradigm to solve the multi-behavior multi-task recommendation problem. Firstly, we simulate the real interaction process of users through the cascading graph paradigm, from auxiliary behaviors (e.g., click, favourite) to the target behavior (i.e., purchase). We learn the embedding of each behavior in a unidirectional manner. Secondly, we introduce a cascading gated feedback module to optimize the representation of the auxiliary behaviors by feeding information from the target behavior to the auxiliary behaviors. Although many gating methods (e.g., MMoE~\cite{MMOE}, PLE~\cite{PLE}) have been proposed, it is not suitable to directly apply these methods to the cascading graph paradigm, because they model a mixture of inputs, which is not suitable to distinguish behavioral preferences. Our cascading gated feedback module allows the cascading graph paradigm to go beyond unidirectional optimization, and explicitly model the effect of the target behavior on the auxiliary behaviors. Thirdly, considering that behavior-specific graphs may lose the higher-order information between users and items, we design a global context enhancement module to further enhance the global consistency of the auxiliary behaviors representation. Previous works~\cite{CRGCN, MBCGCN, AutoDCS} have adopted the cascading graph paradigm, but they do not consider the loss of higher-order information between users and items. Our global context enhancement module constructs a global graph that integrates user interaction data from all behavior layers to provide consistent global preference information to each auxiliary behavior. Finally, to alleviate the user preference bias that may exist in the cascading graph paradigm, we propose a contrastive preference alignment module, which aligns the target behavior embeddings with the global relational embeddings through contrastive learning, thus addressing the preference bias problem that may arise during the multi-behavior cascading process.

We summarize the main contributions as follows:

\begin{itemize}
\item We propose a novel solution called behavior-informed graph embedding learning (BiGEL) for an emerging and important problem, i.e., multi-behavior multi-task recommendation (MMR).
\item We design two key new modules, i.e., cascading gated feedback (CGF) and global context enhancement (GCE). The former regulates auxiliary behavior preferences through gated feedback from the target behavior, ensuring that the cascading process is not merely a one-way optimization. The latter provides a global view of the auxiliary behaviors, ensuring that the user preferences are not affected by the modeling of separate behavior graphs.
\item We design a new contrastive preference alignment (CPA) module to align the target behavior embeddings with the global embeddings using contrastive learning to correct the potential user preference bias in the unidirectional learning process.
\item We conduct extensive experiments on two real-world e-commerce datasets, demonstrating the effectiveness of our BiGEL.
\end{itemize}

\section{Related Work}
In this section, we categorize the related works into single-behavior recommendation, multi-behavior recommendation, and multi-behavior multi-task recommendation.

\subsection{Single-Behavior Recommendation}
Single-behavior recommendation (SBR) aims to personalize ranked lists of items through a single type of interaction between users and items. In traditional SBR, collaborative filtering (CF)~\cite{gsurvey, gsurvey2} is widely used, which utilizes the interaction data between users and items to learn user preferences and item characteristics. Matrix factorization (MF)~\cite{MF} is one of the most classical CF methods, which captures the implicit preferences of users and implicit features of items by representing the relationship between users and items as a matrix and decomposing it into two low-rank matrices. In addition, Bayesian personalized ranking (BPR)~\cite{BPR} optimizes the recommendation performance based on a further assumption that a user's preference for purchased items is usually higher than that for unpurchased items.

In the past decade, more and more deep learning-based SBR methods have been developed. For example, multilayer perceptron (MLP) and autoencoder (AE) are able to effectively capture the latent relationships between users and items by learning more complex nonlinear mappings. Multi-VAE~\cite{VAE} is a variant of autoencoder (AE), whose main idea is to represent the data as a set of latent variables that conform to a certain predefined prior distribution. By sampling from these learned distributions, the model is able to generate new data samples that can effectively capture the underlying patterns and diversity in the data.

In recent years, graph neural networks (GNNs) have also excelled in the field of single-behavior recommendation. GNN-based recommendation methods can leverage graph-structured data to model complex relationships and dependencies between users and items. For example, NGCF~\cite{NGCF} aims to improve the recommendation accuracy by introducing graph convolution combined with higher-order connections between users and items. Compared with traditional CF methods, GNN is able to efficiently integrate multi-level information through a message passing mechanism to capture more complex user-item interactions. LightGCN~\cite{LightGCN} proposes a simplified graph neural network structure that significantly improves the computational efficiency by removing nonlinear activations and residual connections, while still achieving performance comparable to that of complex models on a number of datasets. However, they all target only a single behavior, while in reality users have more than just one type of behavior, making it difficult to accurately learn user preferences from a single type of behavior.

\subsection{Multi-Behavior Recommendation}
Multi-behavior recommendation (MBR) refers to modeling multiple user behaviors to optimize a single target task (i.e., purchase). Traditional matrix factorization methods set different sampling strategies by considering the priority of the target behavior relative to other behaviors. For example, ABPR~\cite{ABPR}, MF-BPR~\cite{MFBPR} and BPRH~\cite{BPRH} are extensions of BPR for addressing MBR problems. In recent years, many researchers have explored multi-behavior models based on graph convolutional networks (GCNs), which can be classified into two categories based on the way they process behaviors, i.e., parallel approaches and cascading approaches.

A parallel approach captures the dependencies between behaviors by learning the embedding of each behavior separately using different methods. For example, MBGCN~\cite{MBGCN} learns the semantics of different behaviors by propagating items based on different user-item behavior propagation layers. GHCF~\cite{GHCF} models different types of user behaviors using graph convolution operations and optimizes the model through multi-task learning. MB-GMN~\cite{MBGMN} designs a graph meta network to learn the dependencies among different behaviors. VCGAE~\cite{VCGAE} combines GCNs and VAE~\cite{VAE} to model multi-behavior interactions. MBSSL~\cite{MBSSL} introduces a novel self-supervised learning framework that adaptively balances the optimization between a self-supervised task and a primary supervised recommendation task by performing node self-discrimination across inter-behavior and intra-behavior dimensions through hybrid gradient operations. MBGCL~\cite{MBGCL} leverages contrastive learning by comparing the target behavior with other auxiliary behaviors for users and items separately, enhancing the representations of different behaviors by introducing noise. MBRCC~\cite{MBRCC} uses GCNs to capture the embeddings of users and items across all behaviors and formulates three contrastive learning tasks: behavior-level embedding contrastive learning, instance-level embedding contrastive learning, and cluster-level embedding contrastive learning. KHGT~\cite{KHGT} considers the relationship between dynamic features of user interactions and knowledge graphs to capture multi-level dependencies of user behaviors through multi-behavior graph Transformer network and an attentive aggregation schema. MBHGCN~\cite{MB-HGCN} learns user preferences by moving from the global level to the behavior-specific level and applies two different aggregation strategies to aggregate user and item embeddings learned from different behaviors. MuLe~\cite{MULE} learns multi-faceted relationships of behaviors through multi-grained graphs, using the attention mechanism to distinguish the uncertain auxiliary behaviors that could potentially lead to purchases or not. BCIPM~\cite{BCIPM} constructs a global view and behavior-specific subgraphs to model users' multiple behaviors. 

A cascading approach leverages the natural order of user behaviors (e.g., \textit{click} $\succ$ \textit{favourite} $\succ$ \textit{purchase}) to design a cascading paradigm, aiming to capture the dependencies between behaviors. For example, CRGCN~\cite{CRGCN} devises a cascading residual graph structure to explicitly exploit behavior relationships in embedding learning. MB-CGCN~\cite{MBCGCN} learns the embedding of each behavior via LightGCN~\cite{LightGCN} and passes the behavior features along the cascading chain through feature transformation. PKEF~\cite{PKEF} introduces a parallel structure on top of the cascading paradigm and employs projection to enhance the behavior representations. AutoDCS~\cite{AutoDCS} uses a bilateral matching gating mechanism based on MB-CGCN to further capture the dependencies between different behaviors. DA-GCN~\cite{DAGCN} considers the possibility of multiple behavior paths in user interactions to better capture the dependencies between different user behaviors. However, most of these methods focus only on the performance of the target behavior (i.e., purchase) task and ignore the performance w.r.t. the auxiliary behaviors.Behavior

\subsection{Multi-Behavior Multi-Task Recommendation}
Multi-task learning (MTL) leverages useful information contained in multiple related tasks to improve the performance of all tasks~\cite{MTLsurvey,M3oE}. Multi-task models can be divided into two categories, i.e., hard parameter sharing and soft parameter sharing. 

Hard parameter sharing is to share the same parameters across different tasks. The shared-bottom model~\cite{share} utilizes lower layers for embedding learning and then uses specific layers to learn each task. ESMM~\cite{ESMM} and ESM2~\cite{ESM2} assume a sequential model of user interactions. They address the problems of sample selection bias and data sparsity by modeling the entire space and transmitting feature representations. In contrast, the soft parameter sharing model has received wider attention due to its higher flexibility and adaptability to task differences. It fuses information between tasks by means of correlation weights between tasks. MMoE~\cite{MMOE} utilizes softmax gated networks to aggregate experts learned from different tasks. PLE~\cite{PLE} is an improvement on MMoE by proposing a customized gate control (CGC) module to explicitly distinguish between the shared and task-specific experts. Multi-behavior multi-task recommendation (MMR) can be seen as a special MTL scenario. It considers each behavior as a task and uses MTL to improve the performance of each behavior simultaneously. Recently, to address the MMR problem, BVAE~\cite{BVAE} extends VAE~\cite{VAE} by incorporating softmax gated aggregate sharing and behavior-specific encoders, and employs task-dependent uncertainty weighted loss (UWL)~\cite{UWL}. POGCN~\cite{POGCN} extends LightGCN~\cite{LightGCN} to learn user and item representations by exploiting edge weighting and partial order BPR, while optimizing the performance of multiple behaviors. 

Despite the significant progress achieved, certain limitations still remain. BVAE~\cite{BVAE} uses a parallel approach to learn the representation of each behavior, ignoring the natural sequential relationship between behaviors in reality and struggling to capture higher-order user-item interactions due to the limitations of the VAE structure. POGCN~\cite{POGCN} learns to share user and item embeddings using a unified graph but does not distinguish user embeddings under different behaviors. This makes it difficult to understand the true preferences of different users across different behaviors. These limitations inspire us to propose a novel solution to address the MMR problem. 

\section{Preliminaries}
\subsection{Problem Definition}
 We denote the set of users as \(\mathcal{U} = \{u_1, u_2, \dots, u_N\}\), the set of items as \(\mathcal{I} = \{i_1, i_2, \dots, i_M\}\), and the set of behaviors as \(\mathcal{B} = \{b_1, b_2, \dots, b_K\}\). We use \(\{b_1, b_2, \dots, b_{K-1}\}\) as the auxiliary behaviors (e.g., click, favourite), and \(b_K\) as the target behavior (i.e., purchase). Each behavior \(b_k\) (\(1 \leq k \leq K\)) represents a specific type of interaction between users and items, which is also treated as a specific recommendation task.
We construct a user-item multi-behavior interaction graph \(\mathcal{G} = (\mathcal{V}, \mathcal{E})\), where \(\mathcal{V} = \mathcal{U} \cup \mathcal{I}\) denotes the set of nodes, and \(\mathcal{E}\) is the set of edges representing interactions of \(K\) different types of behaviors. Given a multi-behavior user-item interaction graph \(\mathcal{G}\), our goal is to predict the most likely item \(i\) that a user \(u\) will interact with under each behavior (i.e., each task). Notations and their explanations are listed in Table~\ref{tab:notations}.

\begin{table}[htbp]
	\centering
        \small
	\caption{Notations and their explanations.}
	\label{tab:notations}
	\begin{tabular}{l|l} 
		\hline\hline
		
		$m$ & the number of users\\
		
		$n$ & the number of items\\
		
		$u$ & the user ID \\
    
		$\mathcal{U}$ & the set of all users\\
  
		$i$ & the item ID \\
  
		$\mathcal{I}$ & the set of all items\\
		
		$\mathbf{e}_u^0$ & the initial embedding of user $u$\\
		
		$\mathbf{e}_i^0$ & the initial embedding of item $i$\\

		$\mathbf{e}_u^{(b_k, l)}$ & the embedding of user $u$ at layer $l$ for behavior $b_k$\\
		
		$\mathbf{e}_i^{(b_k, l)}$ & the embedding of item $i$ at layer $l$ for behavior $b_k$\\
        
            $\mathbf{e}_u^{(b_g, l)}$ & the embedding of user $u$ at layer $l$ in the global context enhancement module\\

             $\mathbf{e}_i^{(b_g, l)}$ & the embedding of item $i$ at layer $l$ in the global context enhancement module\\
		
		$\mathcal{N}_u^{b_k}$ & the set of items interacted with by user $u$ under behavior $b_k$\\
		
		$\mathcal{N}_i^{b_k}$ & the set of users who interact with item $i$ under behavior $b_k$\\

		$\mathbf{e}_u^{(g)}$ & the global embedding of user $u$ obtained from the global graph\\

            $\mathbf{e}_i^{(g)}$ & the global embedding of item $i$ obtained from the global graph\\
        
		$\mathbf{e}_u^{(b_k)}$ & the final embedding of user $u$ under behavior $b_k$\\
		
		$\mathbf{e}_i^{(b_k)}$ & the final embedding of item $i$ under behavior $b_k$\\

		$g_u^{(b_k)}$ & the gating mechanism for user $u$ under behavior $b_k$\\

            $g_i^{(b_k)}$ & the gating mechanism for item $i$ under behavior $b_k$\\

            $y_{ui}^{(b_k)}$ & relevance score for user-item pair $(u, i)$ under behavior $b_k$\\
		
		$\mathcal{L}_{b_k}$ & BPR loss for behavior $b_k$\\
		
		$\mathcal{L}_{cl}$ & total contrastive learning loss\\
		
		$\tau$ & temperature coefficient for softmax function in contrastive learning\\

            $\mathcal{L}$ & overall loss function\\
		
		$\lambda$ & regularization coefficient for contrastive loss\\
		
		$\Theta$ & all trainable parameters\\
		\hline\hline
	\end{tabular}
\end{table}

\section{Methodology}
In this section, we provide a detailed description of our BiGEL. First, we present an illustration of the complete framework. Then, we describe each part of our BiGEL in detail, including cascading GCN learning (CGL), cascading gated feedback (CGF), global context enhancement (GCE), and contrastive preference alignment (CPA). Finally, we discuss the relationship between our BiGEL and the closely related methods.

\begin{figure*}[htbp]
    \centering 
    \includegraphics[width=1\linewidth]{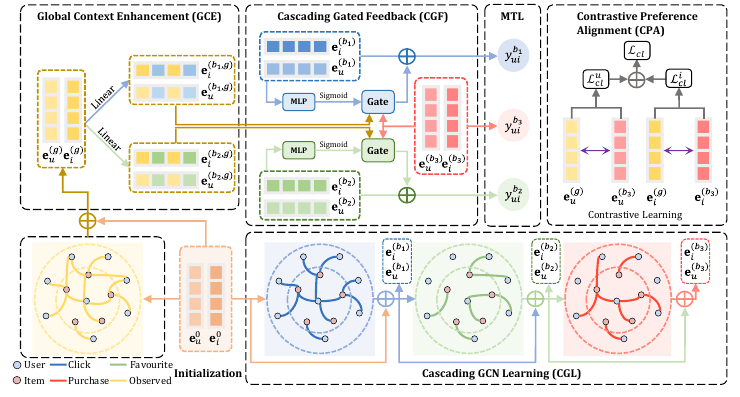} 
    \caption{Illustration of our BiGEL with three user behavior types as an example, i.e., click (\( b_1 \)), favourite (\( b_2 \)), and purchase (\( b_3 \)). First, the cascading GCN learning (CGL) component captures the natural dependencies among different behaviors and models their sequential relationships. Then, the cascading gated feedback (CGF) module uses a gating mechanism to propagate the embedding of the target behavior (\( b_3 \)) back to the preceding auxiliary behaviors (\( b_1 \), \( b_2 \)), and the global context enhancement (GCE) module is designed to refine these behaviors. Next, the contrastive projection alignment (CPA) module aligns the target behavior embedding with the global embedding. Finally, we predict a user's preference scores w.r.t. each behavior type.} 
    \label{fig:framework} 
\end{figure*}

\subsection{Overview}
We illustrate our BiGEL in Figure~\ref{fig:framework}. The CGL component serves initial embedding learning, which enables information transfer from auxiliary behaviors (e.g., click, favourite) to the target behavior (i.e., purchase). Based on CGL, the CGF module selectively feeds back the embedding information of the target behavior to the preceding auxiliary behavior layers through a gating mechanism, enhancing the node representations of these preceding behaviors. The GCE module provides a global graph view to further refine the representations of the auxiliary behaviors. The CPA module corrects the potential user preference bias that may occur during the cascading process by aligning the target behavior embedding with the global embedding. Finally, the whole model is optimized via joint training with multiple loss functions.

\subsection{Embedding Initialization}
Following most GCN-based recommendation models, we first initialize embeddings for each user \( u \in \mathcal{U} \) and each item \( i \in \mathcal{I} \) by transforming their unique one-hot IDs into low-dimensional dense vectors. Specifically, given a user-item pair \((u, i)\), the initial embeddings for user \(u\) and item \(i\) are computed as follows:
\begin{equation}
\mathbf{e}_u^0 = \mathbf{P} \cdot \text{ID}_u^{\mathcal{U}}, \quad \mathbf{e}_i^0 = \mathbf{Q} \cdot \text{ID}_i^{\mathcal{I}},
\label{initial}
\end{equation}
where \(\text{ID}_u^{\mathcal{U}}\) and \(\text{ID}_i^{\mathcal{I}}\) denote the one-hot vectors corresponding to user \(u\) and item \(i\), respectively. \(\mathbf{P} \in \mathbb{R}^{M \times d}\) and \(\mathbf{Q} \in \mathbb{R}^{N \times d}\) are the embedding matrices for users and items, and \(d\) is the embedding dimension.
\subsection{Cascading GCN Learning}
The cascading paradigm has been widely validated as an effective approach for capturing dependencies across different behaviors. We use LightGCN~\cite{LightGCN} as the GCN convolution method due to its notable computational efficiency and demonstrated effectiveness in recommender systems. The embedding propagation process for a user \( u \) and an item \( i \) under behavior \( b_k \) at layer \( l \) is defined as follows:
\begin{equation} 
\mathbf{e}_u^{(b_k, l)} = \sum_{i \in \mathcal{N}_u^{b_k}} \frac{1}{\sqrt{|\mathcal{N}_u^{b_k}|} \sqrt{|\mathcal{N}_i^{b_k}|}} \mathbf{e}_i^{(b_k, l-1)},
\label{eq:global_graph_convu}
\end{equation}
\begin{equation}
\mathbf{e}_i^{(b_k, l)} = \sum_{u \in \mathcal{N}_i^{b_k}} \frac{1}{\sqrt{|\mathcal{N}_u^{b_k}|} \sqrt{|\mathcal{N}_i^{b_k}|}} \mathbf{e}_u^{(b_k, l-1)},
\label{eq:global_graph_convi}
\end{equation}
where \(\mathcal{N}_u^{b_k}\) denotes the set of items that user \( u \) interacts with under behavior \( b_k \), and \(\mathcal{N}_i^{b_k}\) denotes the set of users who interact with item \( i \) under behavior \( b_k \). The term \( \frac{1}{\sqrt{|\mathcal{N}_u^{b_k}|} \sqrt{|\mathcal{N}_i^{b_k}|}} \) is a normalization factor. \(\mathbf{e}_u^{(b_k, l)}\) and \(\mathbf{e}_i^{(b_k, l)}\) denote the embeddings of user \( u \) and item \( i \) after the \( l \)-th layer propagation under behavior \( b_k \), respectively. At layer \( l = 0 \), the embeddings are initialized as \(\mathbf{e}_u^{(b_k, 0)} = \mathbf{e}_u^0\) and \(\mathbf{e}_i^{(b_k, 0)} = \mathbf{e}_i^0\). We denote the final embeddings as \(\mathbf{e}_u^{(b_k)}\) and \(\mathbf{e}_i^{(b_k)}\), which are obtained from the embeddings at the last layer, i.e., \(\mathbf{e}_u^{(b_k)} = \mathbf{e}_u^{(b_k, l)}\) and \(\mathbf{e}_i^{(b_k)} = \mathbf{e}_i^{(b_k, l)}\).

Moreover, following CRGCN~\cite{CRGCN}, we employ a residual connection across the behavior layers to progressively accumulate the corresponding information:
\begin{equation}
\mathbf{e}_u^{(b_k)} = \begin{cases} 
      \mathbf{e}_u^0 + \text{norm}(\mathbf{e}_u^{(b_k)}) & \text{if } k = 1, \\ 
      \mathbf{e}_u^{(b_{k-1})} + \text{norm}(\mathbf{e}_u^{(b_k)}) & \text{if } k > 1, 
   \end{cases}
   \label{eq:crgcn}
\end{equation}
where \(\text{norm}(\cdot)\) represents \(L_2\) normalization. Similarly, we can obtain the embeddings \(\mathbf{e}_i^{(b_k)}\) for each item $i$.

The residual design in Eq.(4) ensures stable information propagation across behavior layers, allowing preferences from sparse auxiliary behaviors (e.g., click) to effectively influence the modeling of downstream behaviors (e.g., purchase). However, a potential drawback is that it may also propagate and amplify noise or bias from preceding layers, particularly when intermediate behaviors are missing for a user. This accumulated bias can distort the user's true preference representation, a phenomenon we refer to as user preference bias. Our subsequent contrastive preference alignment (CPA) in Section~\ref{sec:cpa} is specifically designed to counteract this inherent side effect of the cascading residual structure.


\subsection{Cascading Gated Feedback}
To refine the embeddings of auxiliary behaviors using the target behavior information, we design a cascading gated feedback (CGF) module. While the cascading process in the base model effectively aggregates the information from the auxiliary behaviors to improve the target behavior, it neglects the potential positive impact that the target behavior embeddings can have on refining the auxiliary behaviors. Target behavior embeddings are typically the most explicit expression of user preferences, making them a reliable source for refining the auxiliary behaviors. CGF addresses this limitation by selectively feeding back the target behavior information (i.e., purchase) to the preceding auxiliary behaviors (e.g., click, favourite), enabling a bidirectional and balanced optimization.

Through the CGL process, we obtain the embedding set of user \(u\) across all behaviors as \(\{\mathbf{e}_u^{(b_1)}, \mathbf{e}_u^{(b_2)}, \dots, \mathbf{e}_u^{(b_K)}\}\). To refine the embedding \(\mathbf{e}_u^{(b_k)}\), we integrate the feedback from the target behavior embedding \(\mathbf{e}_u^{(b_K)}\) using a gating mechanism. Unlike the general-purpose gating networks in MMoE~\cite{MMOE} and PLE~\cite{PLE}, which take mixed behavior representations as input and produce coarse-grained gating values with dimension $m$ (i.e., the number of experts), our gating network is specifically designed for a cascading structure. It accepts separate behavior-specific embeddings as input, allowing fine-grained, behavior-wise modulation with dimension $d$ (i.e., the embedding size). Specifically, it applies a linear transformation and a non-linear activation to \(\mathbf{e}_u^{(b_k)}\), followed by a gating mechanism with a sigmoid function to regulate the feedback strength from the target behavior:
\begin{equation}
\begin{aligned}
h_{u}^{(b_k)} &= \text{LeakyReLU}(\mathbf{W}_1^{(b_k)} \mathbf{e}_u^{(b_k)} + \mathbf{b}_1^{(b_k)}), \\
g_{u}^{(b_k)} &= \sigma(\mathbf{W}_2^{(b_k)} h_{u}^{(b_k)} + \mathbf{b}_2^{(b_k)}),
\end{aligned}
\label{eq:gate}
\end{equation}
where \(\mathbf{W}_1^{(b_k)}, \mathbf{W}_2^{(b_k)} \in \mathbb{R}^{d \times d}\), \(\mathbf{b}_1^{(b_k)}, \mathbf{b}_2^{(b_k)} \in \mathbb{R}^d\) are learnable parameters, and \(\sigma(\cdot)\) is the sigmoid function.
With the gating mechanism for each auxiliary behavior \( b_k (k < K) \), we obtain the feedback-enhanced embedding as follows:
\begin{equation}
\mathbf{e}_u^{(b_k)} = \mathbf{e}_u^{(b_k)} + g_u^{(b_k)} \otimes \mathbf{e}_u^{(b_K)},
\label{eq:gate2}
\end{equation}
where $\otimes$ is the element-wise product. Similarly, we can obtain the embeddings \(\mathbf{e}_i^{(b_k)}\) for each item $i$.

\subsection{Global Context Enhancement}
 In the previous subsection, the CGF module feeds back the embedding information of the target behavior (i.e., purchase) to the preceding auxiliary behavior layers.

Since the behavior subgraph convolution alone relies only on the interaction data in the subgraph, it may not fully reflect a user's overall preferences. This is especially true for the auxiliary behaviors, which are inherently contingent or noisy, making it difficult to accurately learn user preferences.

To address this issue, the global context enhancement (GCE) module integrates user interaction data across all behavior layers via a global graph to provide consistent global preference information to the auxiliary behaviors layers. The global graph can also be thought of as a shared layer that aggregates higher-order information across behaviors and distributes it back to optimize the representations of the auxiliary behaviors. Specifically, we treat different behavior as a unified interaction type and use classical GCN to capture high-order cross-behavior information. The convolutional process is defined as follows:
\begin{equation}
\mathbf{e}_u^{(b_g, l)} = \sum_{i \in \mathcal{N}_u} \frac{1}{\sqrt{|\mathcal{N}_u|} \sqrt{|\mathcal{N}_i|}} \mathbf{e}_i^{(b_g, l-1)},
\label{eq:g1}
\end{equation}
where \( b_g \) represents all types of behaviors, \(\mathcal{N}_u\) denotes the set of items that user \( u \) has interacted with, and \(\mathcal{N}_i\) represents the set of users who have interacted with item \( i \). The initial embeddings for the global graph convolution are set as \(\mathbf{e}_u^{(b_g, 0)} = \mathbf{e}_u^0\) and \(\mathbf{e}_i^{(b_g, 0)} = \mathbf{e}_i^0\), which are obtained from the embeddings at the last layer, i.e., \(\mathbf{e}_u^{(g)} = \mathbf{e}_u^{(b_g,l)}\). 

To incorporate global context into auxiliary behavior representation, we first apply a linear transformation to learn behavior-specific global preferences for each auxiliary behavior \( b_k \)\((k < K)\):
\begin{equation}
\hat{\mathbf{e}}_{u}^{(b_k,g)} = \mathbf{W}_3^{(b_k)} \mathbf{e}_u^{(g)} + \mathbf{b}_3^{(b_k)},
\label{eq:g2}
\end{equation}
where \(\mathbf{e}_u^{g}\) denotes the global user representation obtained from the unified graph, and \(\mathbf{W}_3^{(b_k)}\) is a learnable transformation matrix that adapts the global representation to each specific auxiliary behavior. We then integrate the behavior-specific global preferences into each auxiliary behavior representation by applying a gating mechanism:
\begin{equation}
\mathbf{e}_u^{(b_k)} = \mathbf{e}_u^{(b_k)} + g_u^{(b_k)} \otimes \hat{\mathbf{e}}_{u}^{(b_k,g)},
\label{eq:g3}
\end{equation}
where $\otimes$ is the element-wise product. Similarly, we can obtain the embeddings for items \(\mathbf{e}_i^{(b_k)}\). Notably, we do not integrate the global graph information into the target behavior embeddings. This is because the target behavior embeddings are derived through the cascading GCN learning process, which inherently incorporates the information from all preceding behaviors. Hence, direct integration of the global graph information may introduce redundant or even conflicting information, which will interfere with the user preference representation learned through the cascading GCN learning process.

\subsection{Contrastive Preference Alignment}
\label{sec:cpa}
In the first two parts of our BiGEL, we utilize CGF and GCE to optimize the representation of the auxiliary behaviors from different perspectives, making them closer to the true user preferences. However, in real applications, the phenomenon of user preference bias may still occur. This situation is very common in real applications. For example, on an e-commerce platform, user A may click items frequently, but skip the \textit{favourite} behavior before proceeding to purchase. Although the \textit{favourite} behavior layer is missing for user A in the cascading graph model, user A’s preferences are still influenced by user B, who interacts with the same items. Specifically, the \textit{favourite} layer will still propagate interaction information from user B through the convolution operation, which affects user A's preferences and also the subsequent behavior layers. This leads to the phenomenon of user preference bias, as the user interests that should have been reflected by the \textit{favourite} behavior layer are replaced by other users' interaction information. In addition, due to the residual connection of CGL, when some intermediate behaviors (e.g., favourite) are missing, the impact of the \textit{click} behavior will accumulate and pass to a downstream behavior layer, which further exacerbates user preference bias.

To alleviate user preference bias, we design a contrastive preference alignment (CPA) module, which aligns user preferences in the global graph with those of the target behavior through contrast learning. The reason we only perform contrastive learning on the target behavior is that it is a direct reflection of the user's final decision, which can accurately reflect the user preferences. Moreover, since the CGF module is already able to feed back the target behavior to each auxiliary behavior layer for dynamic adjustment, it obviates the need for additional contrastive learning on auxiliary behaviors. We employ the InfoNCE loss~\cite{InfoNCE1, InfoNCE2} to learn from the positive and negative pairs. The user-side contrastive loss is formulated as follows:
\begin{equation}
\mathcal{L}_{cl}^{(u)} = - \sum_{u \in \mathcal{U}} \log \frac{\exp(\Phi(\mathbf{e}_u^{(b_K)}, \mathbf{e}_u^{(g)}) / \tau)}{\sum_{v \in \mathcal{U}} \exp(\Phi(\mathbf{e}_u^{(b_K)}, \mathbf{e}_{v}^{(g)}) / \tau)},
\label{eq:ssl1}
\end{equation}
where \(\Phi(\cdot, \cdot)\) is the cosine similarity function, \(\{(\mathbf{e}_u^{(b_K)}, \mathbf{e}_u^{(g)}) \mid u \in \mathcal{U}\}\) represents positive pairs, and \(\{(\mathbf{e}_u^{(b_K)}, \mathbf{e}_v^{(g)}) \mid u, v \in \mathcal{U}, u \neq v\}\) are negative one. Note that \(b_K\) denotes the target behavior, and \(g\) represents the global behavior aggregation. \(\tau\) is the temperature coefficient used in the softmax function to control the concentration of similarities.

Similarly, we define the item-side contrastive learning loss \(\mathcal{L}_{cl}^{(i)}\), aligning the item's global embedding with its target behavior embedding. By combining the contrastive learning losses for users and items, we have the overall contrastive learning loss:
\begin{equation}
\mathcal{L}_{cl} = \mathcal{L}_{cl}^{(u)} + \mathcal{L}_{cl}^{(i)}.
\label{eq:ssl2}
\end{equation}

\subsection{Multi-Task Learning}

After learning each behavior, we can obtain each user’s embedding set \(\{\mathbf{e}_u^{(b_1)}, \mathbf{e}_u^{(b_2)}, \dots, \mathbf{e}_u^{(b_K)}\}\) and each item’s embedding set \(\{\mathbf{e}_i^{(b_1)}, \mathbf{e}_i^{(b_2)}, \dots, \mathbf{e}_i^{(b_K)}\}\), where \(K\) denotes the number of behavior types. The relevance score \( y_{ui}^{(b_k)} \) for each behavior \( b_k \) (\( 1 \leq k \leq K \)) is computed as the inner product of the embeddings of user \( u \) and item \( i \) as follows:
\begin{equation}
y_{ui}^{b_k} = \mathbf{e}_u^{(b_k)^\mathrm{T}} \mathbf{e}_i^{(b_k)}.
\end{equation}

To optimize the relevance score for each behavior, we adopt the Bayesian personalized ranking (BPR) loss~\cite{BPR}, which is effective in modeling implicit feedback. The BPR loss encourages observed user-item interactions to be scored higher than that of the unobserved ones. The BPR loss for behavior $b_k$ is defined as follows:
\begin{equation}
\mathcal{L}_{b_k} = - \sum_{(u, i, j) \in \mathcal{O}} \ln \sigma(y_{ui}^{b_k} - y_{uj}^{b_k}),
\label{eq:bpr}
\end{equation}
where \( \mathcal{O} = \{(u, i, j) \mid (u, i) \in \mathcal{R}^+, (u, j) \in \mathcal{R}^-\} \) represents the set of positive and negative sample pairs for behavior \( b_k \). Specifically, \(\mathcal{R}^+\) (or \(\mathcal{R}^-\)) denotes the interactions that have been observed (or unobserved) under behavior \( b_k \). Here, \(\sigma(\cdot)\) is the sigmoid function.

Multi-task learning provides a framework for simultaneously optimizing multiple behavior-specific tasks, which enables comprehensive representation learning across different user behaviors. We combine the loss of each task (i.e., each behavior type), and treat each one equally to ensure balanced optimization across different behaviors. The final objective function is formulated as follows:
\begin{equation}
\mathcal{L} = \sum_{k=1}^{K} \mathcal{L}_{b_k} + \lambda \mathcal{L}_{cl} + \beta \|\Theta\|_2,
\label{eq:loss}
\end{equation}
where \(\Theta\) represents the learnable parameters, and \(\lambda\) and \(\beta\) are the weighting coefficients on the contrastive learning loss and the regularization term, respectively.

\subsection{Algorithm Description}
We describe the training process of our BiGEL in the algorithm \ref{alg:BiGEL}. Given the user-item multi-behavior interaction graph \(\mathcal{G}\) and the parameters \(\tau\), \(\lambda\), and \(\beta\), the output is the top-K recommendation results for each user.
The training process consists of several stages. First, we initialize the user and item embeddings \(\mathbf{e}_u^0\) and \(\mathbf{e}_i^0\). Then, we perform LightGCN propagation for each behavior \(b_k\) through the cascading graph paradigm, and compute the embeddings \(\mathbf{e}_u^{(b_k)}\) and \(\mathbf{e}_i^{(b_k)}\). We then compute the gating weights and use feedback from the target behavior \(b_K\) to refine the auxiliary behaviors \(b_k\), integrate the global graph \(\mathbf{e}_u^{(g)}\) with each auxiliary behavior \(b_k\), compute the contrastive preference alignment loss \(\mathcal{L}_{cl}\), and optimize each behavior embedding with the BPR loss \(\mathcal{L}_{b_k}\). Finally, we jointly optimize the loss \(\mathcal{L}_{cl}\) and the losses \(\mathcal{L}_{b_k}\).

\subsection{Complexity Analysis}
Our BiGEL introduces learnable parameters in user and item embedding initialization, as well as in cascading gated feedback (CGF) and global context enhancement (GCE). Specifically, the initial user and item embeddings contribute \(O((M+N)d)\) parameters, where \(M\) and \(N\) are the number of users and items, respectively, and \(d\) is the embedding dimension. CGF and GCE introduce additional parameters for their gating networks and linear transformations, contributing \(O(Kd^2)\) parameters, where \(K\) is the number of behavior types. Thus, the number of total learnable parameters of our BiGEL is \(O((M+N)d + Kd^2)\). The computing complexity of our model is as follows. The time cost is mainly from the cascading graph convolution operations, the operations within the CGF and GCE modules, the contrastive preference alignment (CPA) loss, and the BPR loss. Notice that $|E|$ denotes the total number of edges in the user-item multi-behavior interaction graph across all behaviors, \(n\) is the number of epochs, \(b\) denotes the size of each training batch, and \(L\) represents the number of GCN layers. In the cascading graph convolution process, the computational complexity is \(O(2|E|nKLd/b)\). In the CGF operations, the computational complexity is \(O(n|E|(K-1)d^2/b)\). In the GCE process, the computational complexity is \(O(2|E|nLd/b + n|E|(K-1)d^2/b)\). The computational complexity of the CPA loss is \(O(n|E|bd)\). The computational complexity of the BPR loss across all behaviors is \(O(2|E|nKd)\). Since there are only 1-4 GCN layers and the number of behaviors in multi-behavior tasks is usually very small (\(K = 4\) in UB, and \(K = 3\) in JD), the total computing complexity of our BiGEL is approximately \(O(2|E|nKLd/b + n|E|Kd^2/b + n|E|bd + 2|E|nKd)\). It is worth noting that unlike CRGCN which processes each behavior separately (requiring \(B\) separate computations), our BiGEL processes all behaviors in a unified manner through a single forward pass, making it more computationally efficient for multi-behavior scenarios. Compared with CRGCN, which has a complexity of \(O(2|E|B + 2|E|nBd + 2|E|nBd)\) and requires separate computation for each behavior type (resulting in a total complexity of \(O(2|E|B^2 + 2|E|nB^2d)\) when processing all behaviors), our BiGEL achieves a more efficient one-pass computation that processes all behaviors simultaneously.

\begin{algorithm}[ht]
    \caption{The training process of BiGEL}
    \label{alg:BiGEL}
    \renewcommand{\algorithmicrequire}{\textbf{Input:}}
    \renewcommand{\algorithmicensure}{\textbf{Output:}}
    
    \begin{algorithmic}[1]
        \REQUIRE user-item multi-behavior interaction graph \(\mathcal{G}\), hyperparameters \(\tau, \lambda, \beta\)
        \ENSURE The user top-K recommendation results
        
        \STATE Initialize embeddings \(\mathbf{e}_u^0, \mathbf{e}_i^0\), through Equation~\eqref{initial}
        
        \FOR{$t = 1$ \TO $T$}
            \FOR{$k = 1$ \TO $K$}
                \STATE Perform LightGCN propagation on each behavior \(b_k\), obtain the final user embedding \(\mathbf{e}_u^{(b_k)}\) and item embedding \(\mathbf{e}_i^{(b_k)}\), through Equations~\eqref{eq:global_graph_convu}, \eqref{eq:global_graph_convi}, and \eqref{eq:crgcn}
            \ENDFOR
            \FOR{$k = 1$ \TO $K - 1$}
                \STATE Compute the gating weights for each auxiliary behavior \(b_k\), through Equation~\eqref{eq:gate}
                \STATE Integrate feedback from the target behavior \(b_K\) to refine \(b_k\), through Equation~\eqref{eq:gate2}
            \ENDFOR
            \FOR{$k = 1$ \TO $K - 1$}
                \STATE The auxiliary behaviors \(b_k\) integrates the global graph \(\mathbf{e}_u^{(g)}\), through Equations~\eqref{eq:g1}–\eqref{eq:g3}
            \ENDFOR
            \STATE Compute contrastive preference alignment loss \(\mathcal{L}_{cl}\), through Equations~\eqref{eq:ssl1}–\eqref{eq:ssl2}
            \FOR{$k = 1$ \TO $K$}
                \STATE Compute each behavior embedding BPR loss \(\mathcal{L}_{b_k}\), through Equation~\eqref{eq:bpr}
            \ENDFOR
            \STATE Jointly optimize \(\mathcal{L}_{b_k}\) and \(\mathcal{L}_{cl}\), through Equation~\eqref{eq:loss}
            \STATE Update \(\Theta\), \(\mathbf{e}_u^0\), \(\mathbf{e}_i^0\).
        \ENDFOR
    \end{algorithmic}
\end{algorithm}

\subsection{Discussion}
In this subsection, we discuss the distinctions and connections between our work and the closely related works.

\textbf{CRGCN}.
Our BiGEL is related to CRGCN~\cite{CRGCN}, but is with significant differences. CRGCN uses a unidirectional cascading process to model behavior dependencies, ignoring the positive effect of the target behavior on the auxiliary behaviors and the influence of user preference bias. Our BiGEL fuses feedback from the target behavior and global context in addressing the user preference bias, effectively improving the recommendation performance.

\textbf{BVAE}.
BVAE~\cite{BVAE} learns each behavior representation independently through a VAE encoder and integrates both the global and behavior-specific features using a gating mechanism. However, limited by the fact that the VAE structure does not have the ability to capture high-order interactions between users and items, parallel behavior learning is not able to capture cross-behavior dependencies well. Our BiGEL utilizes GCNs to model high-order interactions between users and items, and employs a cascading process to model the dependencies between different behaviors more accurately.

\textbf{POGCN}.
POGCN~\cite{POGCN} assigns different weights to different behaviors in the convolution process of LightGCN and optimizes the model using a partial order BPR loss. However, it uses a single embedding to represent different behavior types, which limits its ability to capture different interests and preferences that users may exhibit in different behaviors. In contrast, our BiGEL distinguishes users' preferences across different behaviors, enabling finer-grained representations of the interests for each behavior.

\section{Experiments}
In this section, we first introduce the experimental setup in detail. We then conduct extensive empirical studies aiming at exploring the following five research questions.
\begin{itemize}
	\item\textbf{RQ1}. How does our BiGEL perform compared with the state-of-the-art recommendation methods?
	\item\textbf{RQ2}. How do the key modules in our BiGEL affect its overall performance?
	\item\textbf{RQ3}. How will replacing the CGF module with a different MTL module affect the performance?
	\item\textbf{RQ4}. How do the number and order of behaviors affect the performance of our BiGEL?
        \item\textbf{RQ5}. What is the impact of the hyperparameters on the performance of our BiGEL?
\end{itemize}


\subsection{Datasets and Evaluation Metrics}
To evaluate the effectiveness of our BiGEL, following the settings of the previous works~\cite{VCGAE, BVAE}, we adopt two widely used and publicly available datasets, i.e., \textbf{JD\footnote{\url{https://jdata.jd.com/html/detail.html?id=8}}} and \textbf{UB\footnote{\url{https://tianchi.aliyun.com/dataset/dataDetail?dataId=649}}}.

The details of these two datasets are described below.

\textbf{JD.} This dataset is derived from Jingdong, a famous e-commerce platform in China, and contains information about the interactions between 1,608,707 users and 378,457 items. The dataset contains 33,151,074 click records, 441,966 favourite records and 2,193,489 purchase records.

\textbf{UB.} This dataset was sourced from Taobao and published by Alibaba Tianchi platform in 2021. The dataset contains the interaction data of 987,994 users and 4,162,024 items, i.e., 89,716,264 click records, 2,888,258 favourite records, 5,530,446 cart records and 2,015,839 purchase records.

We use the following steps to process these two datasets:  

(i) For duplicate (user, item, behavior) records, we keep only the earliest interaction records. 

(ii) We remove items with less than $k$ purchases (JD: $k=20$, UB: $k=10$). 

(iii) We remove sessions with fewer than $k$ purchases (JD: $k=5$, UB: $k=5$).   

(iv) We keep the timestamp of each record and sort the records w.r.t. their timestamps. The first 80\% of the timeline is used as the training set, 80\%-90\% as the validation set, and the remaining 10\% as the test set.

(v) We remove the users who only interact with items in the validation or test set but not in the training set.

Notably, to avoid the inherent data leakage issue associated with the leave-one-out method~\cite{cutoff}, we follow some previous works by using a global threshold to divide each dataset into training, validation, and test sets~\cite{critical,llara}.Table~\ref{tab:dataset2} summarizes the statistical details of the processed datasets.

\textbf{Evaluation Protocol.}  
We use four metrics to evaluate the performance of the personalized top-K recommendation results, i.e., precision (Prec@K), recall (Rec@K), normalized discounted cumulative gain (NDCG@K), and hit rate (HR@K). As users tend to focus on the top few recommended items~\cite{eva}, we primarily report results when K = 5.


\begin{table*}[h]
  \caption{Statistics of the processed datasets.}
  \label{tab:dataset2}
  \centering
  \resizebox{\textwidth}{!}{
    \begin{tabular}{ l | ccccc | ccccc }
      \toprule
      Statistics 
      & JD & Density 
      & JD (tr.) & JD (val.) & JD (te.) 
      & UB & Density 
      & UB (tr.) & UB (val.) & UB (te.)\\
      \midrule
      \#Users 
      & 10,690 & $-$
      & 10,690 & 4,831 & 4,225
      & 20,443 & $-$
      & 20,443 & 15,827 & 16,086 \\
      \#Items 
      & 13,465 & $-$
      & 12,804 & 6,466 & 5,635 
      & 30,947 & $-$
      & 30,734 & 21,892 & 21,810 \\
      \midrule
      \#Purchase 
      & 71,872 & 0.50\textperthousand
      & 60,967 & 5,892 & 5,013 
      & 133,708 & 0.21\textperthousand
      & 107,489 & 13,088 & 13,131 \\
      \#View 
      & 253,898 & 1.76\textperthousand
      & 217,977 & 20,013 & 15,908 
      & 631,608 & 1.00\textperthousand
      & 511,020 & 59,862 & 60,726 \\
      \#Favorite 
      & 8,810 & 0.06\textperthousand
      & 7,380 & 738 & 692 
      & 26,856 & 0.04\textperthousand
      & 22,270 & 2,345 & 2,241 \\
      \#Cart 
      & $-$ & $-$
      & $-$ & $-$ & $-$ 
      & 82,745 & 0.13\textperthousand
      & 68,207 & 7,397 & 7,078 \\
      \bottomrule
    \end{tabular}
  }
\end{table*}

\begin{table*}[htbp]
  \centering
  \caption{Recommendation performance of our BiGEL and three different categories of representative baselines against different user behaviors on two datasets. We bold the best results and underline the second best results. The superscript $*$ denotes the $p$-value of the significance test between our BiGEL and the second-best method is $p \leq 0.05$. From the distribution of the data and the importance of the behaviors, we consider click and purchase as the main user behaviors, and mark them in bold. Note that the final reported results are the average of three runs using three different random seeds.}
  \resizebox{\linewidth}{!}{
    \begin{tabular}{c|cl|cc|cccccc|cccc}
    \toprule
    \toprule
    \multicolumn{3}{c|}{Models} & MF-BPR & LightGCN  & GHCF  & CRGCN & MB-CGCN & BCIPM & AutoDCS & DA-GCN & BVAE & POGCN & \textbf{BiGEL\phantom{*}} \\
    \midrule
    \multicolumn{1}{c|}{\multirow{12}{*}[-1.2ex]{{JD}}} & \multicolumn{1}{c}{\multirow{4}{*}{\textbf{\#Purchase}}} 
          & Prec@5   & 0.0292{\tiny$\pm$0.0009} & 0.0304{\tiny$\pm$0.0002} & 0.0360{\tiny$\pm$0.0005} & 0.0349{\tiny$\pm$0.0013} & 0.0322{\tiny$\pm$0.0029} & 0.0280{\tiny$\pm$0.0026} & \underline{0.0375}{\tiny$\pm$0.0005} & 0.0365{\tiny$\pm$0.0035} & 0.0360{\tiny$\pm$0.0007} & 0.0324{\tiny$\pm$0.0005} & \textbf{0.0408*}{\tiny$\pm$0.0017} \\
          &       & Rec@5    & 0.0654{\tiny$\pm$0.0029} & 0.0660{\tiny$\pm$0.0017} & 0.0839{\tiny$\pm$0.0003} & 0.0760{\tiny$\pm$0.0051} & 0.0696{\tiny$\pm$0.0025} & 0.0709{\tiny$\pm$0.0048} & \underline{0.0906}{\tiny$\pm$0.0005} & 0.0788{\tiny$\pm$0.0060} & 0.0872{\tiny$\pm$0.0008} & 0.0715{\tiny$\pm$0.0016} & \textbf{0.0976*}{\tiny$\pm$0.0019} \\
          &       & NDCG@5   & 0.0504{\tiny$\pm$0.0016} & 0.0530{\tiny$\pm$0.0006} & 0.0678{\tiny$\pm$0.0004} & 0.0595{\tiny$\pm$0.0027} & 0.0550{\tiny$\pm$0.0028} & 0.0526{\tiny$\pm$0.0040} & 0.0691{\tiny$\pm$0.0019} & 0.0618{\tiny$\pm$0.0049} & \underline{0.0696}{\tiny$\pm$0.0006} & 0.0571{\tiny$\pm$0.0015} & \textbf{0.0739*}{\tiny$\pm$0.0016} \\
          &       & HR@5     & 0.0998{\tiny$\pm$0.0049} & 0.0993{\tiny$\pm$0.0004} & 0.1162{\tiny$\pm$0.0002} & 0.1127{\tiny$\pm$0.0051} & 0.1040{\tiny$\pm$0.0047} & 0.0999{\tiny$\pm$0.0073} & \underline{0.1303}{\tiny$\pm$0.0037} & 0.1159{\tiny$\pm$0.0102} & 0.1211{\tiny$\pm$0.0026} & 0.1029{\tiny$\pm$0.0026} & \textbf{0.1369*}{\tiny$\pm$0.0041} \\
    \cmidrule{2-14}          
    & \multicolumn{1}{c}{\multirow{4}{*}{\textbf{\#Click}}} 
        & Prec@5   & 0.0306{\tiny$\pm$0.0014} & 0.0327{\tiny$\pm$0.0004} & 0.0326{\tiny$\pm$0.0014} & 0.0302{\tiny$\pm$0.0006} & 0.0071{\tiny$\pm$0.0012} & 0.0197{\tiny$\pm$0.0035} & 0.0127{\tiny$\pm$0.0027} & 0.0211{\tiny$\pm$0.0015} & \underline{0.0347}{\tiny$\pm$0.0003} & 0.0326{\tiny$\pm$0.0003} & \textbf{0.0356\phantom{*}}{\tiny$\pm$0.0008} \\
        &       & Rec@5    & 0.0469{\tiny$\pm$0.0022} & 0.0465{\tiny$\pm$0.0024} & 0.0536{\tiny$\pm$0.0020} & 0.0401{\tiny$\pm$0.0022} & 0.0135{\tiny$\pm$0.0019} & 0.0272{\tiny$\pm$0.0022} & 0.0251{\tiny$\pm$0.0042} & 0.0335{\tiny$\pm$0.0020} & \underline{0.0541}{\tiny$\pm$0.0006} & 0.0466{\tiny$\pm$0.0015} & \textbf{0.0569\phantom{*}}{\tiny$\pm$0.0024} \\
        &       & NDCG@5   & 0.0440{\tiny$\pm$0.0027} & 0.0460{\tiny$\pm$0.0010} & 0.0509{\tiny$\pm$0.0025} & 0.0407{\tiny$\pm$0.0008} & 0.0101{\tiny$\pm$0.0014} & 0.0253{\tiny$\pm$0.0031} & 0.0198{\tiny$\pm$0.0037} & 0.0303{\tiny$\pm$0.0023} & \underline{0.0513}{\tiny$\pm$0.0007} & 0.0464{\tiny$\pm$0.0013} & \textbf{0.0526*}{\tiny$\pm$0.0015} \\
        &       & HR@5     & 0.1140{\tiny$\pm$0.0052} & 0.1169{\tiny$\pm$0.0026} & 0.1180{\tiny$\pm$0.0047} & 0.1069{\tiny$\pm$0.0027} & 0.0299{\tiny$\pm$0.0034} & 0.0618{\tiny$\pm$0.0039} & 0.0484{\tiny$\pm$0.0076} & 0.0781{\tiny$\pm$0.0033} & \underline{0.1270}{\tiny$\pm$0.0014} & 0.1125{\tiny$\pm$0.0014} & \textbf{0.1286*}{\tiny$\pm$0.0016} \\
    \cmidrule{2-14} 
    & \multicolumn{1}{c}{\multirow{4}{*}{\#Favourite}} 
          & Prec@5   & 0.0273{\tiny$\pm$0.0007} & 0.0273{\tiny$\pm$0.0007} & \underline{0.0325}{\tiny$\pm$0.0014} & 0.0289{\tiny$\pm$0.0008} & 0.0169{\tiny$\pm$0.0034} & 0.0205{\tiny$\pm$0.0038} & 0.0278{\tiny$\pm$0.0044} & 0.0286{\tiny$\pm$0.0024} & 0.0319{\tiny$\pm$0.0031} & 0.0282{\tiny$\pm$0.0016} & \textbf{0.0348*}{\tiny$\pm$0.0021} \\
          &       & Rec@5    & 0.0537{\tiny$\pm$0.0035} & 0.0517{\tiny$\pm$0.0043} & \underline{0.0727}{\tiny$\pm$0.0073} & 0.0519{\tiny$\pm$0.0012} & 0.0374{\tiny$\pm$0.0048} & 0.0386{\tiny$\pm$0.0083} & 0.0642{\tiny$\pm$0.0110} & 0.0555{\tiny$\pm$0.0064} & 0.0707{\tiny$\pm$0.0068} & 0.0551{\tiny$\pm$0.0058} & \textbf{0.0774\phantom{*}}{\tiny$\pm$0.0095} \\
          &       & NDCG@5   & 0.0425{\tiny$\pm$0.0003} & 0.0443{\tiny$\pm$0.0019} & \underline{0.0551}{\tiny$\pm$0.0051} & 0.0444{\tiny$\pm$0.0014} & 0.0310{\tiny$\pm$0.0056} & 0.0292{\tiny$\pm$0.0048} & 0.0493{\tiny$\pm$0.0098} & 0.0404{\tiny$\pm$0.0041} & 0.0494{\tiny$\pm$0.0059} & 0.0445{\tiny$\pm$0.0031} & \textbf{0.0559\phantom{*}}{\tiny$\pm$0.0037} \\
          &       & HR@5     & 0.0830{\tiny$\pm$0.0052} & 0.0853{\tiny$\pm$0.0034} & \underline{0.1104}{\tiny$\pm$0.0079} & 0.0979{\tiny$\pm$0.0052} & 0.0648{\tiny$\pm$0.0069} & 0.0728{\tiny$\pm$0.0161} & 0.1024{\tiny$\pm$0.0137} & 0.0910{\tiny$\pm$0.0079} & 0.1047{\tiny$\pm$0.0086} & 0.0887{\tiny$\pm$0.0103} & \textbf{0.1240*}{\tiny$\pm$0.0104} \\
    \midrule
    \midrule
        \multicolumn{1}{c|}{\multirow{16}{*}[-1.8ex]{UB}} & \multicolumn{1}{c}{\multirow{4}{*}{\textbf{\#Purchase}}} 
        & Pre@5 & 0.0021{\tiny$\pm$0.0003} & 0.0022{\tiny$\pm$0.0002} & 0.0078{\tiny$\pm$0.0002} & 0.0149{\tiny$\pm$0.0007} & 0.0155{\tiny$\pm$0.0001} & 0.0113{\tiny$\pm$0.0003} & \underline{0.0157}{\tiny$\pm$0.0004} & 0.0081{\tiny$\pm$0.0004} & 0.0068{\tiny$\pm$0.0002} & 0.0095{\tiny$\pm$0.0002} & \textbf{0.0176*}{\tiny$\pm$0.0003} \\
        &       & Rec@5 & 0.0066{\tiny$\pm$0.0010} & 0.0067{\tiny$\pm$0.0005} & 0.0213{\tiny$\pm$0.0007} & 0.0426{\tiny$\pm$0.0022} & 0.0450{\tiny$\pm$0.0010} & 0.0332{\tiny$\pm$0.0010} & \underline{0.0454}{\tiny$\pm$0.0010} & 0.0236{\tiny$\pm$0.0007} & 0.0183{\tiny$\pm$0.0003} & 0.0273{\tiny$\pm$0.0005} & \textbf{0.0497*}{\tiny$\pm$0.0010} \\
        &       & NDCG@5 & 0.0045{\tiny$\pm$0.0007} & 0.0047{\tiny$\pm$0.0002} & 0.0159{\tiny$\pm$0.0002} & 0.0334{\tiny$\pm$0.0014} & 0.0370{\tiny$\pm$0.0007} & 0.0239{\tiny$\pm$0.0007} & \underline{0.0373}{\tiny$\pm$0.0008} & 0.0198{\tiny$\pm$0.0011} & 0.0140{\tiny$\pm$0.0002} & 0.0217{\tiny$\pm$0.0004} & \textbf{0.0406*}{\tiny$\pm$0.0010} \\
        &        & HR@5 & 0.0100{\tiny$\pm$0.0015} & 0.0104{\tiny$\pm$0.0011} & 0.0356{\tiny$\pm$0.0015} & 0.0651{\tiny$\pm$0.0029} & 0.0674{\tiny$\pm$0.0009} & 0.0501{\tiny$\pm$0.0016} & \underline{0.0682}{\tiny$\pm$0.0012} & 0.0360{\tiny$\pm$0.0017} & 0.0305{\tiny$\pm$0.0013} & 0.0417{\tiny$\pm$0.0007} & \textbf{0.0749*}{\tiny$\pm$0.0012} \\
        \cmidrule{2-14}
        & \multicolumn{1}{c}{\multirow{4}{*}{\textbf{\#Click}}} 
        & Pre@5 & 0.0045{\tiny$\pm$0.0002} & 0.0056{\tiny$\pm$0.0001} & 0.0104{\tiny$\pm$0.0005} & 0.0071{\tiny$\pm$0.0003} & 0.0018{\tiny$\pm$0.0001} & 0.0022{\tiny$\pm$0.0003} & 0.0018{\tiny$\pm$0.0002} & 0.0088{\tiny$\pm$0.0002} & \underline{0.0107}{\tiny$\pm$0.0001} & 0.0044{\tiny$\pm$0.0003} & \textbf{0.0215*}{\tiny$\pm$0.0001} \\
        &       & Rec@5 & 0.0059{\tiny$\pm$0.0002} & 0.0073{\tiny$\pm$0.0001} & 0.0168{\tiny$\pm$0.0008} & 0.0107{\tiny$\pm$0.0004} & 0.0025{\tiny$\pm$0.0001} & 0.0033{\tiny$\pm$0.0007} & 0.0024{\tiny$\pm$0.0004} & \underline{0.0192}{\tiny$\pm$0.0004} & 0.0180{\tiny$\pm$0.0007} & 0.0075{\tiny$\pm$0.0005} & \textbf{0.0432*}{\tiny$\pm$0.0002} \\
        &       & NDCG@5 & 0.0063{\tiny$\pm$0.0000} & 0.0075{\tiny$\pm$0.0001} & 0.0165{\tiny$\pm$0.0010} & 0.0105{\tiny$\pm$0.0003} & 0.0022{\tiny$\pm$0.0001} & 0.0030{\tiny$\pm$0.0004} & 0.0023{\tiny$\pm$0.0003} & \underline{0.0175}{\tiny$\pm$0.0002} & 0.0174{\tiny$\pm$0.0004} & 0.0071{\tiny$\pm$0.0005} & \textbf{0.0398*}{\tiny$\pm$0.0002} \\
        &       & HR@5 & 0.0214{\tiny$\pm$0.0009} & 0.0260{\tiny$\pm$0.0005} & 0.0490{\tiny$\pm$0.0025} & 0.0338{\tiny$\pm$0.0008} & 0.0085{\tiny$\pm$0.0002} & 0.0107{\tiny$\pm$0.0016} & 0.0085{\tiny$\pm$0.0009} & 0.0421{\tiny$\pm$0.0009} & \underline{0.0501}{\tiny$\pm$0.0005} & 0.0206{\tiny$\pm$0.0013} & \textbf{0.1000*}{\tiny$\pm$0.0006} \\
        \cmidrule{2-14}
        & \multicolumn{1}{c}{\multirow{4}{*}{\#Favourite}} 
        & Pre@5 & 0.0016{\tiny$\pm$0.0002} & 0.0022{\tiny$\pm$0.0002} & 0.0038{\tiny$\pm$0.0002} & \underline{0.0048}{\tiny$\pm$0.0004} & 0.0038{\tiny$\pm$0.0003} & 0.0020{\tiny$\pm$0.0005} & 0.0032{\tiny$\pm$0.0017} & 0.0042{\tiny$\pm$0.0004} & 0.0025{\tiny$\pm$0.0003} & 0.0028{\tiny$\pm$0.0002} & \textbf{0.0054*}{\tiny$\pm$0.0005} \\
        &       & Rec@5 & 0.0051{\tiny$\pm$0.0008} & 0.0062{\tiny$\pm$0.0007} & 0.0126{\tiny$\pm$0.0015} & 0.0139{\tiny$\pm$0.0012} & 0.0130{\tiny$\pm$0.0008} & 0.0075{\tiny$\pm$0.0019} & 0.0112{\tiny$\pm$0.0057} & \underline{0.0145}{\tiny$\pm$0.0013} & 0.0078{\tiny$\pm$0.0006} & 0.0092{\tiny$\pm$0.0005} & \textbf{0.0181*}{\tiny$\pm$0.0009} \\
        &       & NDCG@5 & 0.0037{\tiny$\pm$0.0006} & 0.0046{\tiny$\pm$0.0002} & 0.0093{\tiny$\pm$0.0007} & 0.0094{\tiny$\pm$0.0011} & 0.0103{\tiny$\pm$0.0007} & 0.0048{\tiny$\pm$0.0012} & 0.0085{\tiny$\pm$0.0050} & \underline{0.0114}{\tiny$\pm$0.0017} & 0.0052{\tiny$\pm$0.0003} & 0.0068{\tiny$\pm$0.0003} & \textbf{0.0134*}{\tiny$\pm$0.0005} \\
        &        & HR@5 & 0.0079{\tiny$\pm$0.0009} & 0.0104{\tiny$\pm$0.0011} & 0.0189{\tiny$\pm$0.0009} & \underline{0.0228}{\tiny$\pm$0.0012} & 0.0189{\tiny$\pm$0.0016} & 0.0097{\tiny$\pm$0.0025} & 0.0162{\tiny$\pm$0.0084} & 0.0195{\tiny$\pm$0.0019} & 0.0125{\tiny$\pm$0.0012} & 0.0136{\tiny$\pm$0.0009} & \textbf{0.0264*}{\tiny$\pm$0.0023} \\
        \cmidrule{2-14}
        & \multicolumn{1}{c}{\multirow{4}{*}{\#Cart}} 
        & Pre@5 & 0.0019{\tiny$\pm$0.0007} & 0.0022{\tiny$\pm$0.0002} & 0.0039{\tiny$\pm$0.0001} &  \underline{0.0046}{\tiny$\pm$0.0002} & 0.0015{\tiny$\pm$0.0001} & 0.0017{\tiny$\pm$0.0003} & 0.0022{\tiny$\pm$0.0013} & 0.0034{\tiny$\pm$0.0002} & 0.0029{\tiny$\pm$0.0001} & 0.0022{\tiny$\pm$0.0002} & \textbf{0.0050*}{\tiny$\pm$0.0003} \\
        &       & Rec@5 & 0.0052{\tiny$\pm$0.0025} & 0.0066{\tiny$\pm$0.0015} & 0.0112{\tiny$\pm$0.0002} &  \underline{0.0149}{\tiny$\pm$0.0006} & 0.0053{\tiny$\pm$0.0003} & 0.0060{\tiny$\pm$0.0010} & 0.0076{\tiny$\pm$0.0043} & 0.0112{\tiny$\pm$0.0003} & 0.0087{\tiny$\pm$0.0005} & 0.0073{\tiny$\pm$0.0004} & \textbf{0.0166*}{\tiny$\pm$0.0011} \\
        &       & NDCG@5 & 0.0036{\tiny$\pm$0.0019} & 0.0043{\tiny$\pm$0.0011} & 0.0086{\tiny$\pm$0.0001} &  \underline{0.0103}{\tiny$\pm$0.0004} & 0.0032{\tiny$\pm$0.0003} & 0.0036{\tiny$\pm$0.0004} & 0.0052{\tiny$\pm$0.0039} & 0.0080{\tiny$\pm$0.0007} & 0.0067{\tiny$\pm$0.0001} & 0.0051{\tiny$\pm$0.0003} & \textbf{0.0117*}{\tiny$\pm$0.0006} \\
        &       & HR@5 & 0.0092{\tiny$\pm$0.0033} & 0.0100{\tiny$\pm$0.0008} & 0.0182{\tiny$\pm$0.0001} &  \underline{0.0220}{\tiny$\pm$0.0009} & 0.0073{\tiny$\pm$0.0003} & 0.0086{\tiny$\pm$0.0012} & 0.0107{\tiny$\pm$0.0067} & 0.0169{\tiny$\pm$0.0005} & 0.0138{\tiny$\pm$0.0005} & 0.0109{\tiny$\pm$0.0012} & \textbf{0.0243*}{\tiny$\pm$0.0011} \\
    \bottomrule
    \bottomrule
    \end{tabular}%
    }
  \label{tab:Overall}
\end{table*}%

\subsection{Baselines and Parameter Configurations}
\subsubsection{Baselines}
To comprehensively evaluate the performance of our BiGEL, we compare it with ten representative baselines, which can be divided into three groups.

\noindent\textbf{Single-Behavior Recommendation Methods}
\begin{itemize}
    \item \textbf{MF-BPR}~\cite{BPR}: It is based on the BPR strategy, which assumes that positive samples score higher than negatives.
    \item \textbf{LightGCN}~\cite{LightGCN}: This is a simplified GCN model that removes feature transformations and activations, retaining only neighborhood aggregation, which performs quite well among GCN-based recommendation methods.
\end{itemize}

\noindent\textbf{Multi-Behavior Recommendation Methods}
\begin{itemize}
    \item \textbf{GHCF}~\cite{GHCF}: It is a graph neural network model that encodes higher-order heterogeneous collaborative signals using combinatorial operators and introduces an effective non-sampling learning strategy.
    \item \textbf{CRGCN}~\cite{CRGCN}: It adopts a cascading GCN structure in order to model multi-behavior data. The features learned from one behavior are passed to the next behavior through a residual design. Multi-task learning is also used to optimize the model.
    \item \textbf{MB-CGCN}~\cite{MBCGCN}: The model is first pre-trained, and then the cascading dependencies are handled through a chain of LightGCN blocks, where the preceding behaviors are passed directly to the next after feature transformation.
    \item \textbf{BCIPM}~\cite{BCIPM}: The model captures complex relationships between user and item for a given behavior and enhances preference representation with GCNs. It employs joint optimization using a cross-entropy loss and a BPR loss.
    \item \textbf{AutoDCS}~\cite{AutoDCS}: It builds on MB-CGCN by using bilateral matching gating to identify the importance of different behavior types from the user and item perspectives, relaxing the decision chain constraints of the cascading graph paradigm.
    \item \textbf{DA-GCN}~\cite{DAGCN}: This model constructs personalized directed acyclic behavior graphs to capture dependencies, using a GCN-based encoder for message propagation and an attention module to optimize the embeddings. A multi-task learning method adjusts the contribution of each auxiliary behavior.
\end{itemize}

\noindent\textbf{Multi-Behavior Multi-Task Recommendation Methods}
\begin{itemize}
    \item \textbf{BVAE}~\cite{BVAE}: This model extends VAE by dividing the encoder into two parts. It optimizes the model from both global and behavior-specific perspectives. The global interaction and fusion behavior representations are weighted using two gated networks with a loss function adapted to the task weight.
    \item \textbf{POGCN}~\cite{POGCN}: This model is optimized by extending the LightGCN structure into a partial order graph, assigning different weights to the edges, and improving the BPR loss using the behavior combinatorial probability.
\end{itemize}

\subsubsection{Experimental Details.}
We implement our BiGEL in PyTorch\footnote{\url{https://pytorch.org/}}. We set all embedding sizes to 100, batch sizes to 500, and number of training epochs to 1000, and use Adam as the optimizer~\cite{BVAE, VCGAE}. For our BiGEL, the learning rate is searched from $\{0.0001, 0.001, 0.01\}$, the regularization weight $\beta$ is fixed as $1 \times 10^{-3}$, the contrastive learning weight $\lambda$ is searched from $\{0.1, 0.01, 1\}$, and the temperature coefficient $\tau$ is searched from $\{0.1, 0.2, 0.5, 0.8\}$. The number of graph neural network layers $L$ in the global graph is searched from $\{1, 2, 3, 4\}$. For behavior-specific graphs, we adopt a heuristic method to determine the number of layers, increasing the layer count progressively based on the sequential order of the behaviors, as we consider a later behavior more important. Specifically, the number of layers for JD behavior chain (\textit{click} $\succ$ \textit{favourite} $\succ$ \textit{purchase}) is set to $\{1, 2, 3\}$. The UB behavior chain (\textit{click} $\succ$ \textit{favourite} $\succ$ \textit{cart} $\succ$ \textit{purchase}) is set to $\{1, 2, 3, 4\}$. For the other baselines, we use their official open-source code and carefully tune the parameters to find the best performance for comparison. To ensure a fair comparison~\cite{konstan,ferrar}, we conducted rigorous hyperparameter tuning for all methods. Specifically, for all baselines, the learning rate is searched from $\{0.0001, 0.001, 0.01\}$. For the eight GNN-based models, i.e., LightGCN, GHCF, CRGCN, MB-CGCN, BCIPM, AutoDCS, DA-GCN, and POGCN, the number of graph neural network layers $L$ is searched from $\{1, 2, 3, 4\}$. For GHCF, we further search various combinations of behavior-specific weights to find the best configuration. For BCIPM, the loss hyperparameter $\beta$ is searched from $\{0.1, 0.3, 0.5, 0.7, 0.9\}$. For DA-GCN, the trade-off hyperparameter $\alpha$ between the target and auxiliary tasks is searched from $[0.2, 3.0]$ with a step of $0.2$. For POGCN, the temperature coefficients $\tau$ and $\gamma$ are searched from $[0.2, 5.0]$ with a step of $0.2$. All optimal values of the aforementioned parameters are determined according to the performance of NDCG@5 on the validation set.

Note that for single-behavior recommendation methods, we mix all behaviors in the training set without distinguishing behavior types. In both single-behavior recommendation methods and multi-behavior multi-task recommendation methods, the sum of the performance w.r.t. each behavior is used as the early stopping criterion. For multi-behavior recommendation methods, the target behavior performance is used as the early stopping criterion, because they do not perform well if using the same criterion as that of the other two types of methods.

\begin{figure}[t]
    \setlength{\belowcaptionskip}{-0.3pt}   
    \centering
    \begin{subfigure}[b]{0.9\linewidth}
        \centering
        \includegraphics[width=\linewidth]{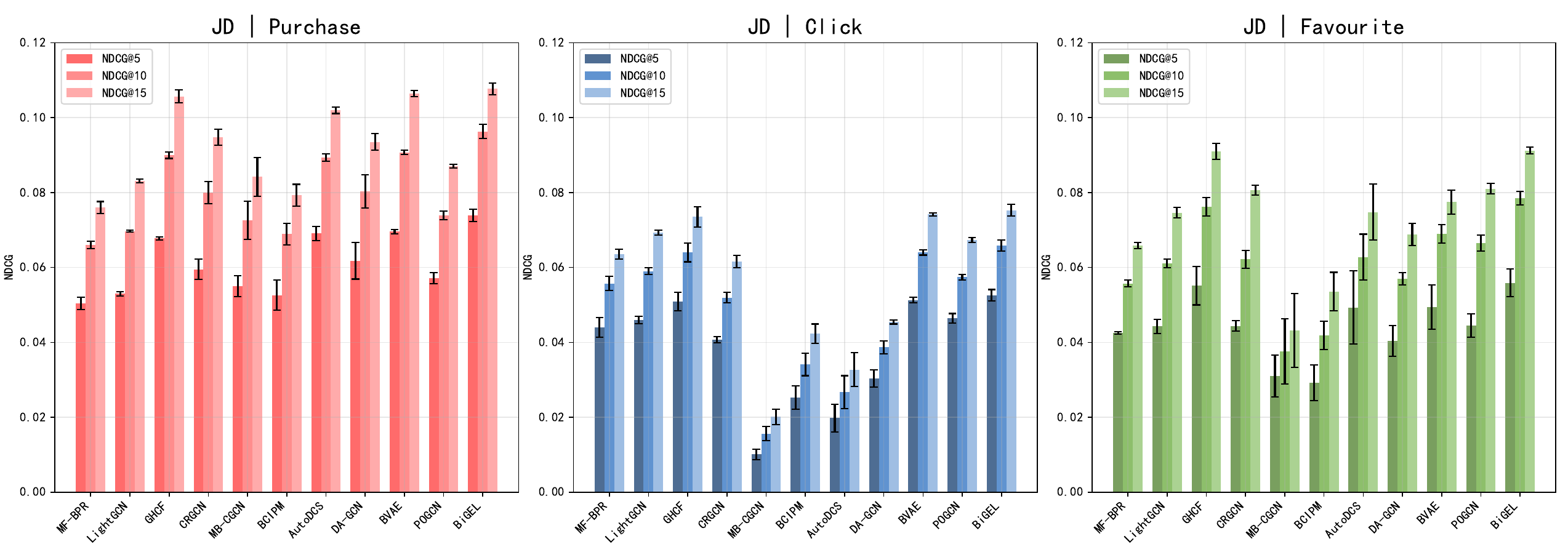}
    \end{subfigure}
    \hfill
    \begin{subfigure}[b]{\linewidth}
        \centering
        \includegraphics[width=\linewidth]{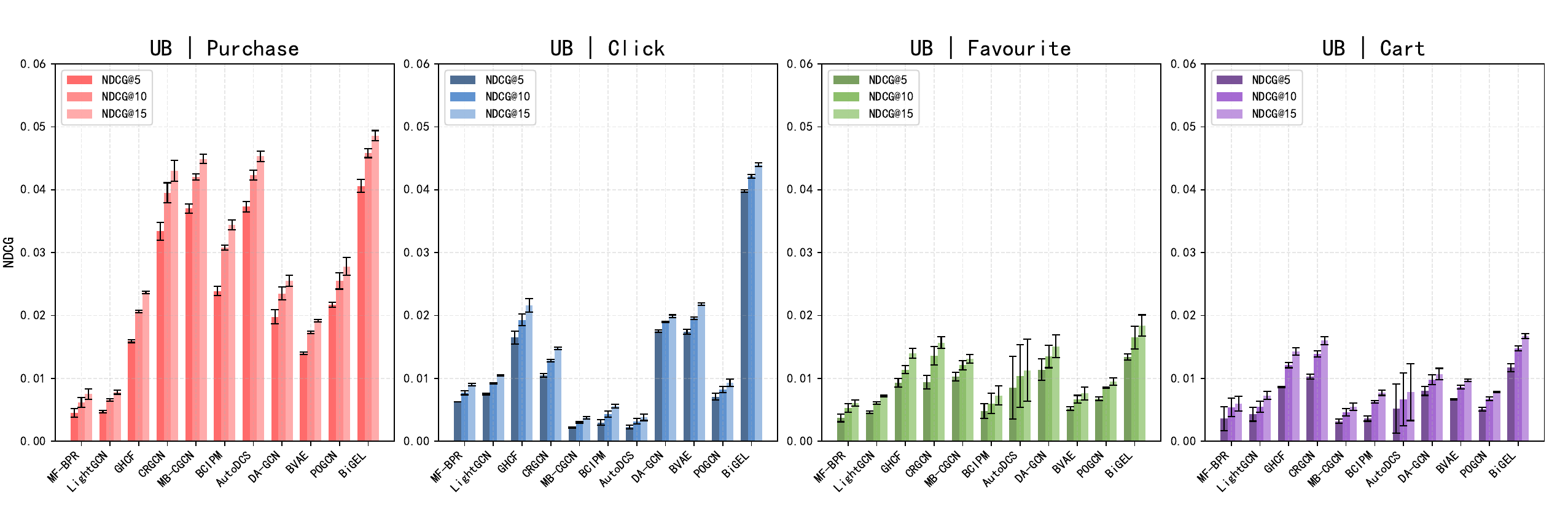}
    \end{subfigure}
    \caption{Recommendation performance on JD (top) and UB (bottom) with different cutoff values K.}
    \label{fig:RQ1}
\end{figure}

\begin{figure}[ht]
    \setlength{\belowcaptionskip}{1pt}   
    \centering
    \begin{subfigure}[b]{0.65\linewidth}
        \centering
        \includegraphics[width=\linewidth]{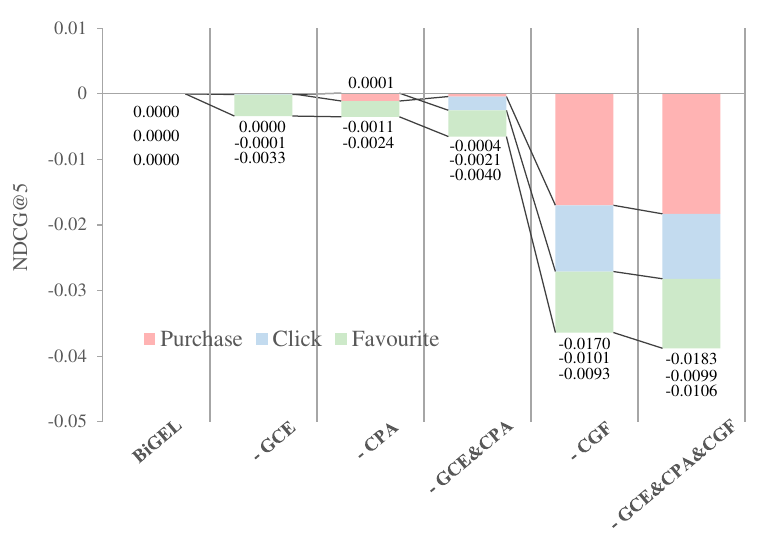}
        \caption{Variation of NDCG@5 relative to BiGEL on JD.}
    \end{subfigure}
    \hfill
    \begin{subfigure}[b]{0.65\linewidth}
        \centering
        \includegraphics[width=\linewidth]{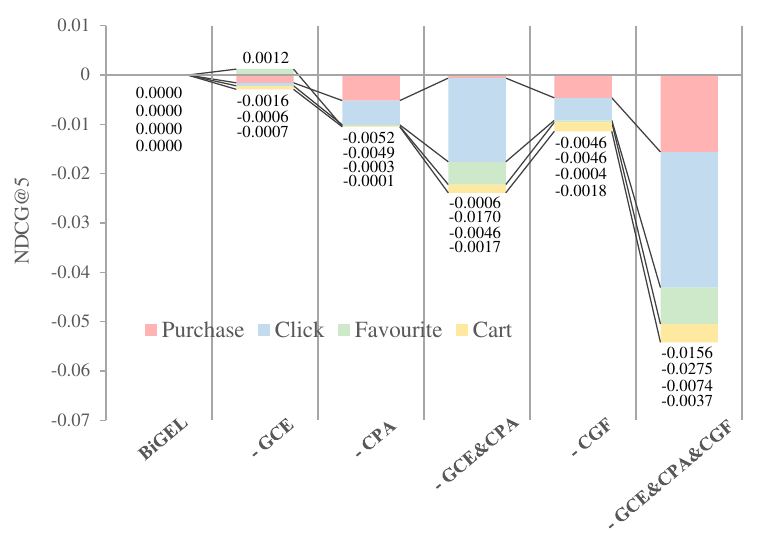}
        \caption{Variation of NDCG@5 relative to BiGEL on UB.}
    \end{subfigure}
    \caption{Variation of the recommendation performance when removing GCE, CPA, CGF or their combinations.}
    \label{fig:RQ2}
\end{figure}

\subsection{Overall Performance (RQ1)}
We report the main results of the baseline methods and our BiGEL in Table~\ref{tab:Overall}. From Table~\ref{tab:Overall}, we can get the following observations: 

(i) In most cases, multi-behavior recommendation methods are better than single-behavior recommendation methods in terms w.r.t. each behavior target behavior, which suggests that it is essential to consider multi-behavior information on recommender systems. 

(ii) Most multi-behavior recommendation methods primarily focus on improving the target behavior while neglecting the performance of auxiliary behaviors. 

(iii) The multi-behavior multi-task recommendation methods BVAE and POGCN show more balanced performance between the target and auxiliary behaviors on JD, especially with significant improvement on the auxiliary behaviors. 

(iv) The cascading graph paradigm-based methods CRGCN, MB-CGCN, and AutoDCS outperform most of the other baseline methods on the target behavior, but not on the auxiliary behaviors, which prompts us to think about how to balance the performance of the target behavior and auxiliary behavior in order to better solve the MMR problem. 

(v) Our BiGEL effectively compensates for the limitations of the cascading graph paradigm by considering the feedback of the target behavior during the cascading process and the addition of auxiliary behavior representations from the global perspective to achieve optimal performance, especially with respect to the most important and primary user behaviors (i.e., purchase and click).

To avoid potentially misleading conclusions from a small cutoff (e.g., @5)~\cite{valcarce2018}, we include more results at higher cutoffs (e.g., @10 and @15) of NDCG. The results on other metrics are similar. As shown in Figure~\ref{fig:RQ1}, the NDCG values of all models increase with a higher cutoff, and the overall trend is consistent with the results in Table~\ref{tab:Overall}.

\subsection{Ablation Study (RQ2)}
In order to analyze the contribution of each module to the overall performance, we perform ablation experiments. Specifically, we consider five variants: 

(1) removal of the global context enhancement module (denoted as "w/o GCE").

(2) removal of the contrastive preference alignment module (denoted as "w/o CPA").

(3) removal of the cascading gated feedback module (denoted as "w/o CGF").

(4) removing the GCE and CPA modules (denoted as "w/o GCE\&CPA").

(5) removing the GCE, CPA and CGF modules (denoted as "w/o GCE\&CPA\&CGF"). 

It is worth noting that in the third variant, the GCE module is also indirectly removed due to the removal of the CGF module. We report the results using the NDCG@5 metric as an example, and all other metrics turn out to be similar. All results are based on the average of three different random seeds and are displayed in Figure~\ref{fig:RQ2}. From the results, we have the following observations:

1) If we remove GCE and CGF, i.e., "w/o GCE" and "w/o CGF", two metrics show a slight performance increase, and most other metrics show a decrease. This is particularly evident in the "w/o GCE \& CPA \& CGF" variants, where the decline is very pronounced.

2) If we remove GCE, i.e., "w/o GCE", there is a decrease in both click and favourite behaviors on JD. On UB, there is a decrease in all behaviors except the favourite behavior. This phenomenon may stem from inconsistencies in the representation of user preferences between the favourite behavior and the other behaviors (i.e., click, cart, and purchase) in UB. The favourite behavior has a relatively weak representation of user intent in UB and accounts for a mere 0.04\% of overall behaviors. Therefore, directly incorporating the global preferences into the sparse behavior may introduce noise.

3) If we remove CGF, i.e., "w/o CGF", there is a significant decrease in all behaviors except the favourite behavior on UB, which indicates that the feedback regulation mechanism of the target behavior is effective for the enhancement of the auxiliary behaviors. It achieves the feedback-driven process of the cascading structure.

4) If we remove CPA, i.e., "w/o CPA", there is a decrease in the purchase behavior on both JD and UB, suggesting that the user preference bias in the cascading process will negatively affect the performance of the model. When the cascading propagation replaces the missing intermediate behaviors (e.g., favourite) with other users’ interaction information, it distorts the true user preferences and introduces bias. CPA alleviates  the user preference bias by aligning the comprehensive user preferences from the global graph with the target behavior representations via contrastive learning.

\begin{figure}[t]
    \centering
    \includegraphics[width=0.85\linewidth]{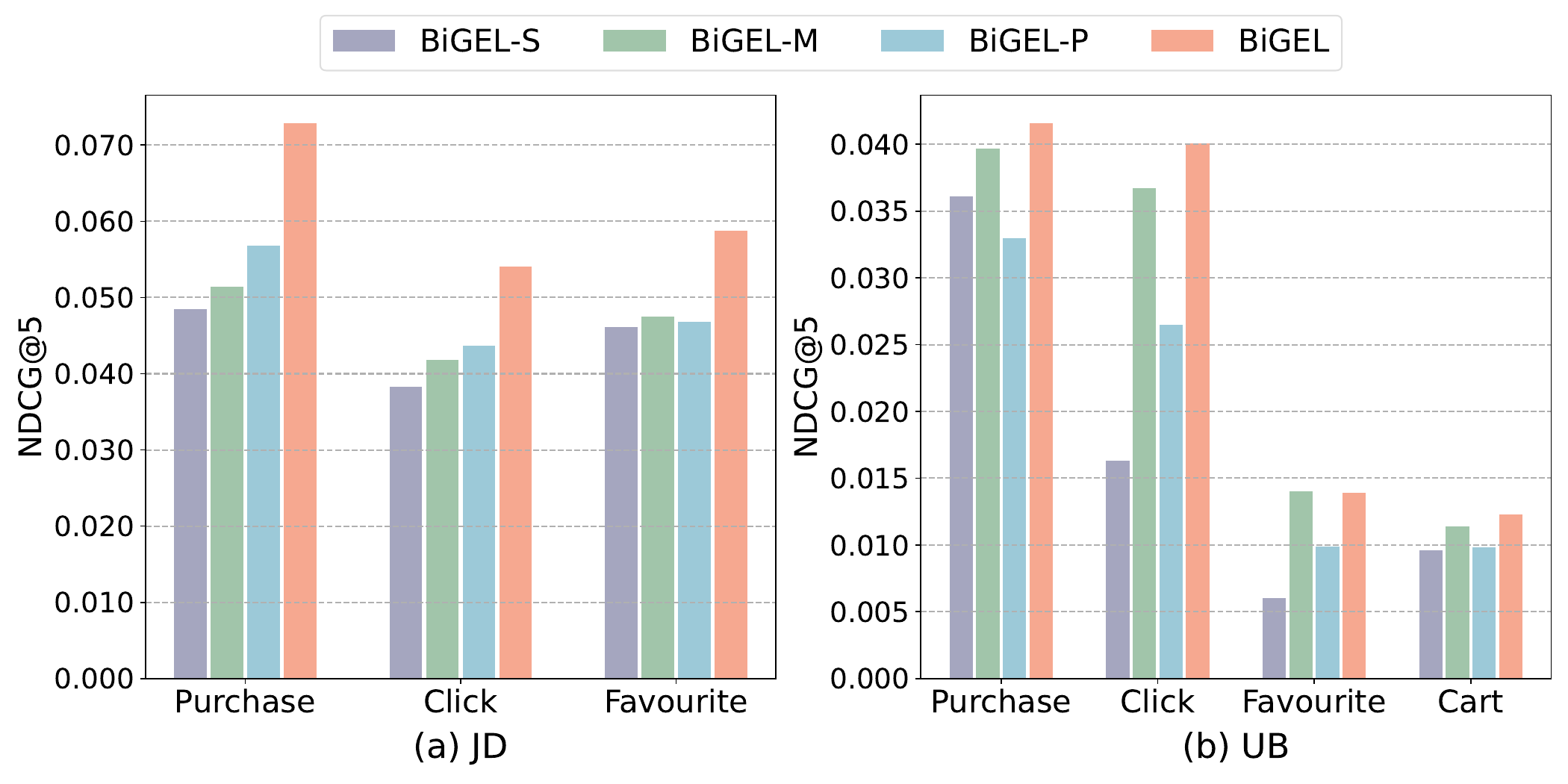} 
    \caption{Impact of different MTL modules on our BiGEL.} 
    \label{fig:RQ3} 
\end{figure}

\subsection{Impact of the MTL Module (RQ3)}
In order to further validate the effectiveness of our proposed CGF module, we conduct module substitution experiments. Specifically, we remove the CGF (including GCE) module, and then use the global graph convolution as a shared expert and each behavior graph convolution as a specific expert. We use three different MTL methods, i.e., Shared-Bottom~\cite{share}, MMoE~\cite{MMOE}, and PLE~\cite{PLE}, which are named, BiGEL-S, BiGEL-M, and BiGEL-P, respectively. The experimental results are shown in Figure~\ref{fig:RQ3}. We can see that our BiGEL works best on all tasks, except that BiGEL-M slightly outperforms our BiGEL on the favourite task on UB. This is likely that the favourite data on UB is too sparse. BiGEL-P outperforms BiGEL-M on JD's purchase and click, and BiGEL-M outperforms BiGEL-P on favourite. For UB, BiGEL-M outperforms BiGEL-P on all behaviors, and BiGEL-S outperforms BiGEL-P on purchase. This is due to the increase in the number of tasks. In BiGEL-P, each task only acquires knowledge from shared experts. As a result, it becomes difficult to capture the complex behavior dependencies. Finally, our BiGEL performs best on the sum of all behavior performance on both datasets.

\subsection{Impact of the Number and Order of Behaviors (RQ4)}
We explore the effect of the number and order of behaviors on our BiGEL. Due to the fact that the favourite (Fav) and cart behaviors of JD and UB are too sparse compared with purchase and click, we focus on the main user behaviors, i.e., purchase and click. 

\begin{figure}[t]
    \captionsetup{skip=1pt} 
    \centering
    \includegraphics[width=1\linewidth]{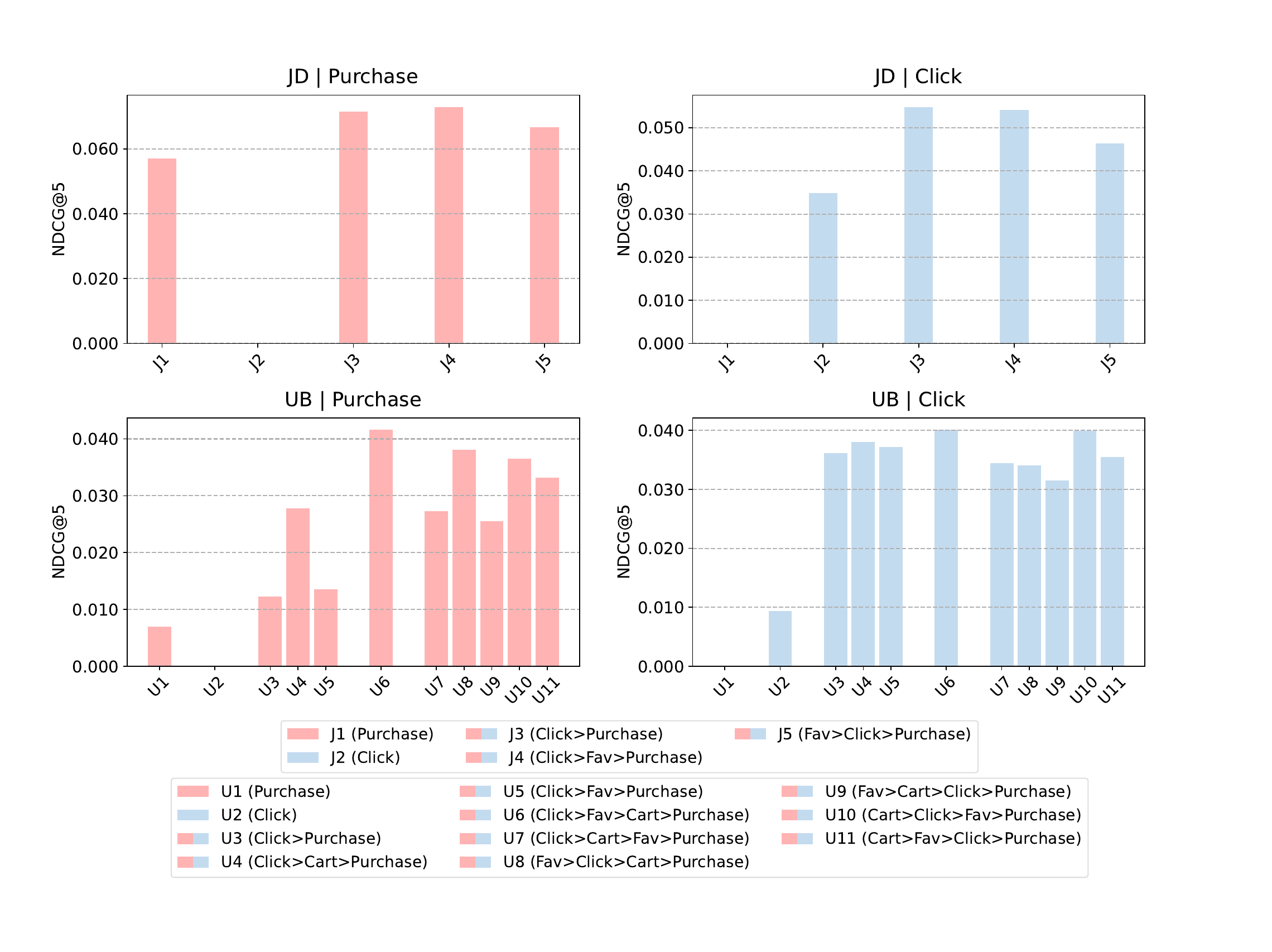} 
    \caption{Recommendation performance of our BiGEL w.r.t. different behaviors (i.e., purchase on the left and click on the right) on JD (top) and UB (bottom).} 
    \vspace{-0.2cm}
    \label{fig:RQ4} 
\end{figure}

The experimental results are shown in Figure~\ref{fig:RQ4}. We have the following observations.

\textbf{Number of behaviors.}
1) For the purchase behavior of JD and UB, the performance gradually improves as the number of behaviors increases. For example, from J1~(\textit{Purchase}), J3~(\textit{Click} $\succ$ \textit{Purchase}) to J5~(\textit{Fav} $\succ$ \textit{Click} $\succ$  \textit{Purchase}) and from U1~(\textit{Purchase}), U3~(\textit{Click} $\succ$ \textit{Purchase}), U4~(\textit{Click} $\succ$ \textit{Cart} $\succ$ \textit{Purchase}), U5~(\textit{Click} $\succ$ \textit{Fav} $\succ$ \textit{Purchase}) to U6~(\textit{Click} $\succ$ \textit{Fav} $\succ$ \textit{Cart} $\succ$ \textit{Purchase}), increasing the number of behaviors provides richer contextual information and enhances task synergy, improving the performance of all behavior prediction tasks.
2) For the click behavior of JD, from J2~(\textit{Click}) to J3~(\textit{Click} $\succ$ \textit{Purchase}), the performance also gradually improves. However, in J4~(\textit{Click} $\succ$ \textit{Fav} $\succ$ \textit{Purchase}), the performance decreases after adding the favourite behavior, probably because the direct correlation between the favourite behavior and the click behavior on JD is weak, and the introduction of the favourite behavior introduces a certain amount of noise to the modeling of the click behavior.
3) For the click behavior of UB, similar to the purchase behavior of UB, the performance also gradually improves with the increase of the number of behaviors.

\textbf{Behavior order.} 
For the purchase and click behaviors of JD and UB, changing the predefined behavior order resulted in a decrease in model performance. For example, the performance of the purchase and click behaviors on JD decreases when the behavior order is changed from J4~(\textit{Click} $\succ$ \textit{Fav} $\succ$ \textit{Purchase}) to J5~(\textit{Fav} $\succ$ \textit{Click} $\succ$  \textit{Purchase}), and the performance of the purchase and click behaviors on UB decreases when the behavior order is changed sequentially from U6~(\textit{Click} $\succ$ \textit{Fav} $\succ$ \textit{Cart} $\succ$ \textit{Purchase}) to U7~(\textit{Click} $\succ$ \textit{Cart} $\succ$ \textit{Fav} $\succ$ \textit{Purchase}), U8~(\textit{Fav} $\succ$ \textit{Click} $\succ$ \textit{Cart} $\succ$ \textit{Purchase}), U9~(\textit{Fav} $\succ$ \textit{Cart} $\succ$ \textit{Click} $\succ$ \textit{Purchase}), U10~(\textit{Cart} $\succ$ \textit{Click} $\succ$ \textit{Fav} $\succ$ \textit{Purchase}), and U11~(\textit{Cart} $\succ$ \textit{Fav} $\succ$ \textit{Click} $\succ$ \textit{Purchase}). This result suggests that incorrect cascade paths usually have some degree of impact on the model performance.

\begin{figure}[tp]
    \captionsetup{skip=-2pt} 
    \centering
    \includegraphics[width=0.75\textwidth]{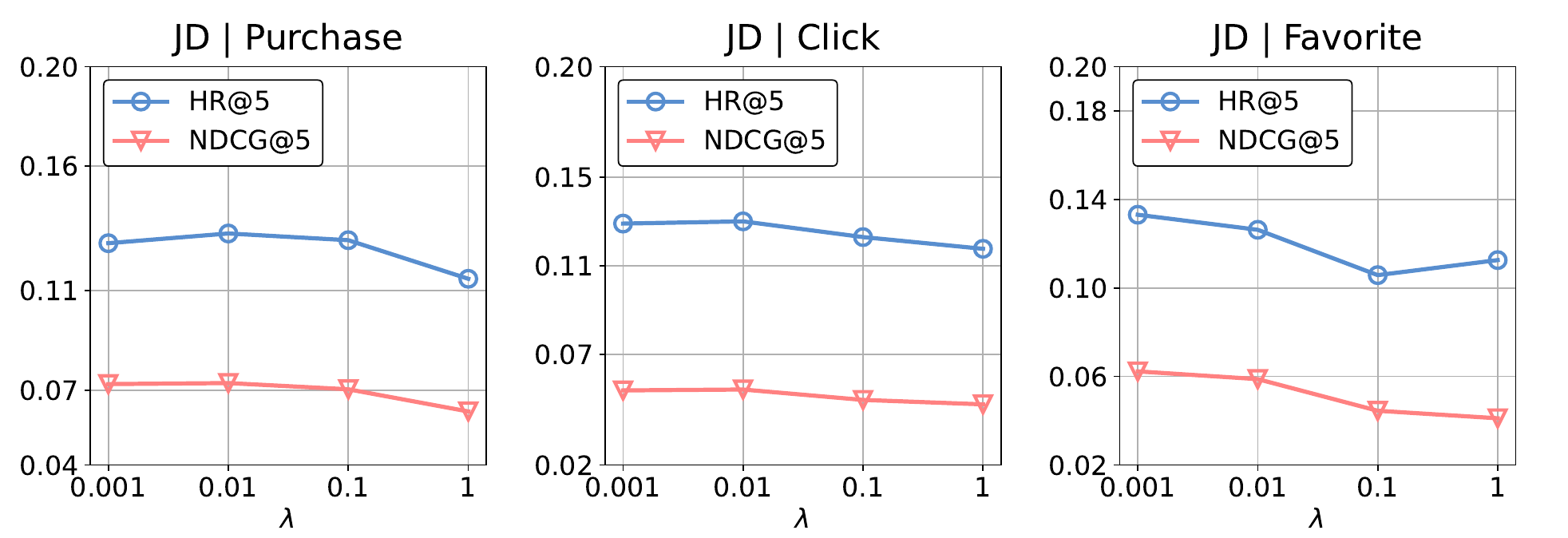} 
    \vspace{0.8cm} 
    \includegraphics[width=\textwidth]{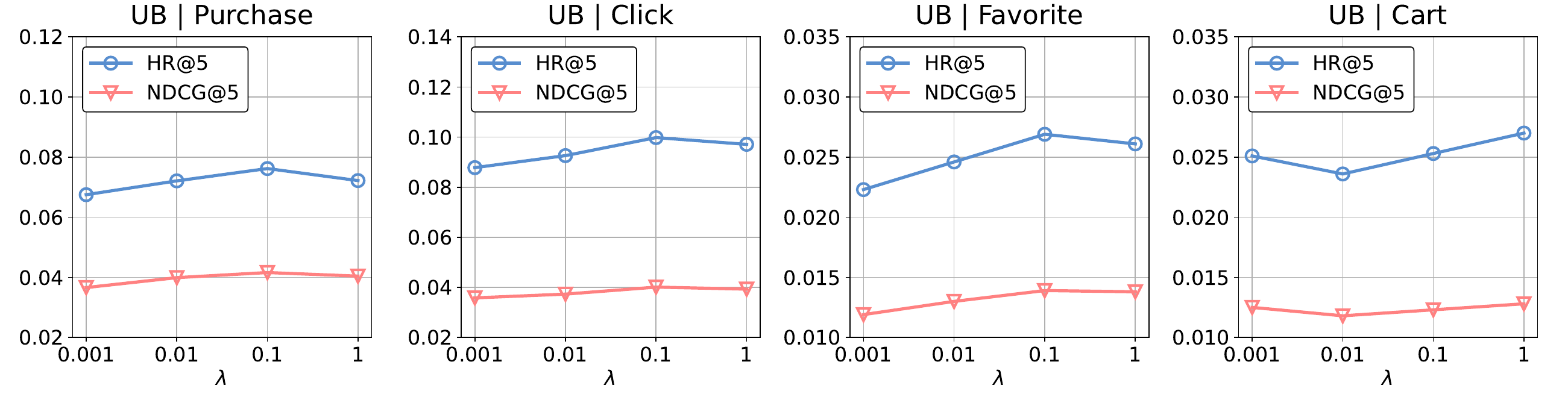} 
    \caption{Recommendation performance on JD (top) and UB (bottom) with different values of contrastive learning weights $\lambda$.}
    
    \label{fig:RQ5}
\end{figure}

\subsection{Hyper-Parameter Analysis (RQ5)}
We explore the effect of contrastive learning on the model's performance by varying the weight of the contrastive learning loss in the range of $\{0.001, 0.01, 0.1, 1\}$. We report the changes on HR@5 and NDCG@5 metrics for the three behaviors of the JD dataset and the four behaviors of the UB dataset for different contrastive learning loss weights, which are shown in Figure~\ref{fig:RQ5}. We have the following observations: For the JD dataset, the best performance for click and purchase behaviors occurs at $\lambda = 0.01$. Performance decreases as $\lambda$ increases, suggesting that too high a contrastive learning loss weight may lead to overfitting or introduce negative bias. However, for the favourite behavior, the best performance is at $\lambda = 0.001$. This may be because the favourite behavior is a small proportion of the JD dataset, and higher contrastive learning weight helps improve its performance. For the UB dataset, the best performance occurs at $\lambda = 0.1$ for click, favourite, and purchase behaviors. Performance decreases when $\lambda$ is increased or decreased. For the cart behavior, the best performance is at $\lambda = 1$. This may be due to the introduction of more noise during the cascading process, requiring a larger $\lambda$ to constrain the preference of the cart behavior. For both the JD and UB datasets, we observe that as the number of behaviors increases, a larger $\lambda$ lead to better performance. This helps alleviate the problem of user preference bias.

\section{Conclusions and Future Work}
In this paper, we propose a novel solution called behavior-informed graph embedding learning (BiGEL) for multi-behavior multi-task recommendation. Our BiGEL contains three embedding learning modules, i.e., GCE, CGF and CPA, on top of behavior-informed graphs. Specifically, each behavior embedding is initially learned via cascading GCN. Then through CGF, the information of the target behavior is fed back to each auxiliary behavior for refinement. Subsequently, GCE is used to integrate the user interaction data from all behavior layers to provide consistent global preference information for each auxiliary behavior. Finally, the target behavior embeddings are aligned with the global embeddings through CPA, where contrastive learning is utilized to correct the potential user preference bias in graph embedding learning. Extensive experiments on two public datasets show that our BiGEL achieves superior performance compared with the state-of-the-art methods.

For future works, we mainly consider two aspects. On the one hand, our BiGEL sets the same weight for each behavior task, while our preliminary studies show that it is largely independent of the weight on the behavior-specific losses. We are thus interested in studying how to exploit the complementary effect of the multi-task network structure in our BiGEL and behavior-specific loss weight learning in GradNorm~\cite{gradnorm}. On the other hand, we will explore the sequential information in BiGEL to capture users' dynamic preferences more accurately.


\bibliographystyle{ACM-Reference-Format}
\bibliography{sample-base}

\appendix

\end{document}